# The Effect of Shear-Thinning Rheology on the Dynamics and Pressure Distribution of a Single Rigid Ellipsoidal Particle in Viscous Fluid Flow

A. Awenlimobor[1], D. E. Smith[1*]

[1]Department of Mechanical Engineering, School of Engineering and Computer Science, Baylor University, Waco, TX 76798, USA

*Corresponding author. Email: douglas_e_smith@baylor.edu

**Abstract**

This paper evaluates the behavior of a single rigid ellipsoidal particle suspended in homogenous viscous flow with a power-law Generalized Newtonian Fluid (GNF) rheology using a custom-built finite element analysis (FEA) simulation. The combined effects of the shear-thinning fluid rheology, the particle aspect ratio, the initial particle orientation and the shear-extensional rate factor in various homogenous flow regimes on the particles dynamics and surface pressure evolution are investigated. The shear-thinning fluid behavior was found to modify the particle's trajectory and alter the particle's kinematic response. Moreover, the pressure distribution over the particle's surface is significantly reduced by the shear-thinning fluid rheology. The FEA model is validated by comparing results of the Newtonian case with results obtained from the well-known Jeffery's analytical model. Furthermore, Jeffery's model is extended to define the particle's trajectory in a special class of homogenous Newtonian flows with combined extension and shear rate components typically found in axisymmetric nozzle flow contractions. The findings provide an improved understanding of key transport phenomenon related to physical processes involving fluid-structure interaction (FSI) such as that which occurs within the flow-field developed during material extrusion-deposition additive manufacturing of fiber reinforced polymeric composites. These results provide insight into important microstructural formations within the print beads.

**Introduction**

Theoretical analysis of particle behavior in a viscous homogenous suspension is a well-known Fluid Structure Interaction (FSI) problem which has a variety of applications in key transport phenomena observed in physical rheological systems such as the movement of cells and platelets in blood plasma[1], the motion of reinforcing particles in fiber-filled polymer melt suspensions during polymer composite processing[2], proppants transport in fracturing fluids[3], migration of gaseous bubbles in quiescent viscous flows[4] etc. The rheology of particle suspensions is inherently complex due to a host of factors, including the presence of inter and intra particle forces arising from hydrodynamic interaction, contact collision between particles, confinement effect and particle deformability, Brownian disturbance, non-Newtonian viscoelastic fluid rheology, anisotropic particle geometry and concentration, and existence of various flow regimes within the system, etc.[5,6,7]. The study of particle suspension dynamics often starts with the evaluation of single rigid spherical particle suspension under Newtonian simple shear flow which also provides insight into the rheology of dilute suspensions[8,9]. As an example, the dynamics of a single rigid ellipsoidal axisymmetric particle has been used extensively to investigate particle dynamics and flow-field structure of polymer composite melt flows during processing to assess their microstructure[7,10,11].

Theoretical studies on particle migration in homogeneous viscous flow are commonly based on the assumptions of negligible inertia effects, Newtonian fluid rheology and non-deformable particle shape, conventionally referred to as "standard conditions" [12]. Pioneering works of Oberbeck[13], Edwardes[14] and Jeffery[15] evaluated the orbit of an ellipsoidal rigid particle suspended in a homogenous shear viscous flow, where particle motion was determined to be a function of initial condition which has been validated experimentally[16]. In other work, Bretherton showed that lateral positioning of spherical isotropic particles remains unchanged relative to their initial position in quiescent sedimentation or unidirectional shear viscous flow[17]. Additionally, Cox found that the orientation of transversely isotropic rigid particles in unconfined quiescent sedimentation would remain fixed at its initial value throughout its motion[18]. These studies showed that under 'standard conditions', the motion and trajectory of a body of revolution depends on its initial conditions. For instance, the so called 'degeneracy' of Jeffery's orbits is used to describe the indeterminacy of particle's motion in sheared viscous suspension whereby an axisymmetric particle may assume any of the infinitely possible metastable periodic orbits depending on its initial position. Experimental observations have revealed a tendency for suspended particles to eventually acquiesce to an equilibrium configuration within a finite timescale or equilibrium rate of approach irrespective of its initial configuration contrary to theoretical predictions based on "standard conditions"[12]. Jeffery[15] first suggested the possibility that spheroidal particles in a sheared viscous suspension with a theoretically indeterminate nature based on first order approximations, may eventually assume special configurations which are the path of least energy dissipation. Taylor[19] was one of the earlier researchers to provide experimental basis for Jeffery's hypothesis and proposed that the higher order terms neglected in Jeffery's approximate equations were responsible for the observed departure in the actual particle's behavior from theoretical predictions. In a separate experimental study Saffmann et al.[20] showed that suspended particle's do not always settle in preferred configuration states, however when they do, the contributions of the non-Newtonian fluid viscosity neglected in Jeffery's approximate equations accounted mainly for the observed discrepancy between theoretical predictions and actual particle's behavior. Other non-linear effects such as fluid and particle inertia, confinement and end effect were found to be infinitesimal as to significantly alter the particle's motion within a finite timescale. Jeffery's equations are generally found to be sufficient in predicting particle's kinematics in a dilute and semi-dilute viscous shear-thinning particle suspension yielding only minor deviations from experimentally observed response[21,22]. However, in the concentrated regime, the Jeffery's model are no longer valid in predicting particle's motion as the departures of theoretical prediction from experimental observations becomes significant due to the combined effect of short range fiber interactions and shear-thinning fluid rheology neglected in Jeffery's model assumptions[21]. The effect of other rheological properties on the dynamics of a suspended particle such as higher order viscoelasticity fluid behavior that may be found in actual FSI physical systems have also been investigated by several researchers. An increase in the fluid elasticity results in a slow drift of prolate spheroids in sheared viscous suspension across spectrum of degenerate Jeffery orbits from a tumbling orbit to a log-rolling state and at drift rates proportional to the shear rate[23,24]. Moreover, an excessive shear rate was found to result in particle realignment with the prevailing flow direction and the critical shear rate for flow realignment depended on the particles aspect ratio and Ericksen's number.

Computational models that account for particle inertia, non-Newtonian fluid rheology and/or shape deformability have more recently emerged. These more advanced models are often used to assess the departure of each from related theoretical predictions of fiber kinematics based on standard conditions. These advanced models are either developed from analytically formulations based on variational principles or asymptotic series expansion about the limits of standard theoretical model assumptions[12] or developed from numerical based simulations[25]. Analytical models are relatively faster, and computationally more efficient compared to numerical models, however these models are non-flexible, often restricted to predicting unique quantities and are less accurate due to oversimplification[26]. Methods based on variational principle are used to define limit bounds on the hydrodynamic drag coefficient of a spherical particle in GNF fluid subject to creeping flow[25]. The method has been successfully applied to obtain limit bounds solutions on the drag for spheres in GNF fluids for different viscosity models including the Newtonian model[27], power-law model[28], the Carreau model[29] and the Ellis model[30]. The approach is more accurate for predicting hydrodynamic bounds in just Newtonian and power-law fluid models and the limit bounds diverges with increasing shear-thinning[25]. Perturbation based methods are generally used to compute solutions of fluid flow at relatively low Weissenberg number[31]. For instance, asymptotic perturbation about the leading order Newtonian fluid model has been used to evaluate the motion of transversely isotropic rigid particles in second-order viscoelastic fluid suspension[31,32]. Consistent with experimental observations, at low shear rates, the viscoelastic fluids cause the suspended particle to slowly drift through various Jeffery's orbit until the attainment of an equilibrium orientation state in the flow vorticity direction. At higher shear rates, particles re-orient with the flow direction and its rotation suppressed. Extension of the theory to particle shapes revealed that while prolate spheroids tend towards a log-rolling position in the vorticity direction, oblate spheroids had an affinity for tumbling in the flow plane[33]. On the contrary, application of the perturbation technique to investigate the effect of weakly shear-thinning fluid rheology on particles motion in unconfined sheared viscous suspension revealed that the degeneracy of Jeffery's orbit where unaffected by the non-Newtonian fluid rheology[34]. However, Jeffery's orbit and period were found to be instantaneously modified by the shear-thinning fluid behavior, and the quantitative modifications depended on the particle's initial conditions.

Prior research that utilized numerical simulation techniques are summarized in various review literature[6, 35,36,37]. The method is tenable to increased model complexity and improved idealization of actual physical systems with increased accuracy. However, the method is computationally intensive and suffers from high computational cost. Numerical based models are classified into mesh-free or particle-based methods (PBM) and the traditional gridded continuum or element-based method (EBM)[35,36]. To avoid detraction from the primary focus of this paper, the reader is referred to existing review literature for more details[35,36]. PBM is a meshless, fully Lagrangian-based highly adaptive technique that allows for instantaneous tracking of individual particle response within a heterogenous multiphase system and capable of modeling flow fronts, free surfaces and accurately solving large deformation problems[38-42]. Examples of PBM includes the explicit Smoothed Particle Hydrodynamic (SPH) and the Moving Particle Semi-Implicit (MPS) method and Discrete Element Method (DEM). Although PBM has been applied to evaluate the development of complex single-phase flows with non-linear fluid rheology[43-46], the behavior of suspended particles in non-linear suspension flow are seldom evaluated with this method. Typical DEM solution techniques include the Dynamic Numerical Simulation (DNS), Lattice Boltzmann Method (LBM), and particle Finite Element Analysis (pFEA). Applications of DEM to FSI problems are summarized in various literature[6, 37]. DEM has been extensively used to study the behavior of single particles in Newtonian viscous suspension[47-51] and also in non-linear viscous suspension[52, 53].

EBM types includes the Finite Element Method (FEM), the Finite Difference Method (FDM), the Finite Volume Method (FVM) and the Boundary Element Method (BEM)[35,37]. In EBM, individual domain units are interconnected via topological maps. EBM involves transformation of a complex Partial Differential Equation (PDE) into a system of repetitive but simplified algebraic equations with solutions computed at the unit nodes, cells or elements level and collated to yield an approximate general solution. EBM is a well-established and highly evolved numerical technique extensively utilized in solving Computational Fluid Dynamics (CFD) and FSI transport problems. For single particle suspension, an extra physical modelling that involves balancing the net hydrodynamic forces and couples on the surface of the particle is required to compute the particles motion. FDM and FVM has been used to compute flow field and fiber orientation dynamics in mold filling process[54-56]. BEM has been successfully implemented to study flow-field development of particulate suspension in viscous shear flow[57-59] and FEM has been used to study single particle behavior in linear viscous shear flow[60-63]. Relevant to this study are the applications of EBM in non-linear single particle suspension. For instance, 2D FEM has been used to simulate single rigid spheroidal particle behavior in dilute non-linear viscous shear flow[25,62,64]. The findings revealed that shear-thinning effect only slightly affected the particle's kinematic, and this impact diminishes with increasing fiber slenderness. Moreover, increased shear-thinning was found to significantly reduced the magnitude of the pressure distribution surrounding the particle surface although had negligible effect on the shape pressure profile itself[25, 64].

By reduction from 3D to 2D space numerical techniques have also been used to study the effect of dimensional space on flow-field response surrounding a particle based on Jeffery's model[63]. While the particle's motion was observed to be unaffected by the dimensional space, the pressure distribution was found to differ significantly. Although a 2D analysis may suffice to study particle's motion in viscous suspension, a 3D analysis is necessary to accurately predict the particle's surface pressure distribution. It is evident that extensive literature on the behavior of axisymmetric particles in viscous suspension exists, however, previous studies have focused on the evolution of particles dynamics and are mostly based on linear shear flow. The local flow field structure surrounding the particle, including the velocity and pressure distribution is seldom investigated which are particularly relevant in understanding complex processes involved in physical rheological systems. Moreover, existing studies that also investigated development of the pressure field surrounding a particle are based on flow analysis around fixed particle in space[25,53,65] that do not consider the influence of the particle's dynamics on the pressure distribution.

The present study utilizes 3D FEM based simulation to investigate the effect of non-standard Jeffery's condition including the effect of generalized Newtonian fluid (GNF) rheology on the dynamics and surface pressure distribution of a single particle suspended in viscous homogenous flows. Firstly, we explore the effect of various factors such as the fibers geometric aspect ratio and initial fiber angle on the single particles motion and surface pressure distribution for a single particle suspended in Newtonian homogenous flow-field using Jeffery's equation. Typical size of particles encountered during Extrusion Deposition Additive Manufacturing (EDAM) polymer composite processing are on average hundreds of microns in magnitude depending on the particles concentration and system's scale, usually around $50 - 100 \mu m$ for small scale EDAM systems and $\sim 300 \mu m$ for large scale EDAM systems[66]. The rotary Peclet number that characterizes these polymeric melt flow through an EDAM nozzle are orders of magnitude high (i.e. $Pe_r \gg 1$). Brownian effects arising from particle interaction with the surrounding fluid molecules are thus insignificant and have been ignored in the current investigation since the hydrodynamic forces are expected to dominate the particle's motion. Jeffery's equations are a good starting point for studying particles behavior in these Newtonian flows. More rigorous stochastic statistical analysis accounting for Brownian disturbance such as that conducted by Leal et al.[67] and Zhang et al.[68] is a relevant study for future consideration. The generalized Newtonian FEA single fiber motion model development is a non-linear extension to the Newtonian formulations of Zhang et al. [68-70] and Awenlimobor et al.[63,64] assuming a power-law non-Newtonian fluid behavior for fiber suspension rheology. A two (2) stage Newton Raphson numerical algorithm is used in our simulation, firstly to solve for the steady-state flow-field velocities and pressure distribution within the flow domain and secondly to compute the resulting translational and rotational velocities of the rigid spheroidal particle during its motion in various homogenous flow fields by equilibrating the net force and couple acting on the particles surface and the fiber's instantaneous positions and orientations are updated using a numerical ordinary differential equation (ODE) solution technique. FEA model validation is achieved by comparing steady state responses at a single time step of the quasi-transient analysis of a single particle motion along Jeffery's orbit obtained from a custom-built FEA simulation with results obtained Jeffery's Equations. Likewise, the behavior of the particle (kinematics and surface pressure response) in various Newtonian homogenous flow fields are benchmarked for both Jeffery's Model and FEA simulation. Finally, we investigate the resulting effect of particle shape and the shear-thinning fluid rheology on the particle's dynamics and evolution of the pressure distribution response on the fibers' surface in the various homogenous flow fields using our validated FEA model. These findings are particularly useful in controlling process parameters to optimize the microstructure of particulate polymer composites to improve print properties.

## Methodology

This section provides in detail the methods used for predicting the behavior of a single three-dimensional (3D) rigid ellipsoidal particle suspended in Newtonian and non-Newtonian viscous homogenous shear-extension flows. The first section presents Jeffery's formulation for the flow-field development around an ellipsoid and explicit derivations for the particle motion (angular velocities and orientation angles) in a special class of linear homogenous flow with combined extension and shear rate velocity gradient components that idealizes typical flow conditions found in various sections of an EDAM extruder-nozzle. The second section details the FEA model development for obtaining particle angular velocities, orientation angles and field velocities and pressure distribution surrounding a particle suspended in non-linear creeping shear flow with a power-law fluid definition. Subsequent section presents results of the model validation by comparing the evolution of the particle's angular velocities and surface pressure distribution obtained from both Jeffery's analytical equations and FEA numerical model for different Newtonian flow cases and particle aspect ratio. Except stated otherwise, we consider a geometric aspect ratio of $r_e = 6$ for the prolate spheroid, a consistency index of $m = 1\ Pa \cdot s^n$ for the power-law fluid or a viscosity of $\mu_1 = 1\ Pa \cdot s$ for Newtonian fluid, and shear rate of $\dot{\gamma} = 1\ s^{-1}$ for the various flow cases.

### **Standard Jeffery Analytical Model**

Jeffery[15] derived analytical equations for the motion of a single 3D ellipsoidal particle suspended in a Newtonian homogenous viscous creeping flow by linearization of the Navier Stokes equations assuming a zero Reynolds number. The following includes a summary of Jeffery's particle-fluid interaction dynamics model where he obtained expressions for the velocity and pressure field within the fluid surrounding the particle. The equations for the pressure and velocity within a Newtonian fluid having viscosity $\mu_1$ are respectively given as

$$p = p_0 + 2\mu_1 \Lambda_{ij}^{III} \nabla_i \nabla_j \Omega \qquad 1$$

and

$$\dot{X}_i = \dot{X}_i^\infty + \nabla_i \Lambda_j^I \chi_j + \epsilon_{ijk} \nabla_j \Lambda_{km}^{II} X_m + \Lambda_{jk}^{III} X_k \nabla_i \nabla_j \Omega - \Lambda_{ij}^{III} \nabla_j \Omega \qquad 2$$

where the position vector $\underline{X}$, gradient operator $\underline{\nabla}$ and integral function $\underline{\chi}$ are given respectively as

$$\underline{X} = [X_1 \quad X_2 \quad X_3]^T, \qquad \underline{\nabla} = \left[\frac{\partial}{\partial X_1} \quad \frac{\partial}{\partial X_2} \quad \frac{\partial}{\partial X_3}\right]^T, \qquad \underline{\chi} = [\chi_1 \quad \chi_2 \quad \chi_3]^T \qquad 3$$

In the above, the Laplace function $\Omega$ is defined in terms of the independent position vector variables $\underline{X}$ and $\lambda$ as

$$\Omega = \Omega(\underline{X},\lambda) = \int_\lambda^\infty \frac{1}{\Delta}\left\{\frac{X_1^2}{a^2+\lambda} + \frac{X_2^2}{b^2+\lambda} + \frac{X_3^2}{c^2+\lambda} - 1\right\} d\lambda, \qquad \Delta = \{(a^2+\lambda)(b^2+\lambda)(c^2+\lambda)\}^{1/2} \qquad 4$$

where $\lambda$ is an arbitrary offset distance from the particle's surface obtained from the positive real roots of

$$\frac{X_1^2}{a^2+\lambda} + \frac{X_2^2}{b^2+\lambda} + \frac{X_3^2}{c^2+\lambda} = 1, \qquad \lambda \geq 0 \qquad 5$$

The undisturbed fluid velocity $\dot{X}_i^\infty$ in eqn. 2 above is given as

$$\dot{X}_i^\infty = L_{ij} X_j \qquad 6$$

where $L_{ij}$ is the velocity gradient tensor. The constant-coefficient tensors $\Lambda_i^I, \Lambda_{ij}^{II}$ & $\Lambda_{ij}^{III}$ that appear in eqns. 1 and 2 above are given as

$$\underline{\Lambda}^I = \begin{bmatrix} R \\ S \\ T \end{bmatrix}, \qquad \underline{\underline{\Lambda}}^{II} = \begin{bmatrix} U & & \\ & V & \\ & & W \end{bmatrix}, \qquad \underline{\underline{\Lambda}}^{III} = \begin{bmatrix} A & H & G' \\ H' & B & F \\ G & F' & C \end{bmatrix} \qquad 7$$

where expressions for the components shown here are given in Appendix A. The terms in $\Lambda_{ij}^{III}$ are simply the stresslet and torque acting on the rigid ellipsoid suspended in linear ambient flow-field[71]. The tensors $\Lambda_i^I, \Lambda_{ij}^{II}$ & $\Lambda_{ij}^{III}$ are functions of the symmetric rate of deformation tensor $\Gamma_{ij}$ and the antisymmetric vorticity tensor $\Xi_{ij} = \epsilon_{imn}\Xi_m\delta_{nj}$ obtained by decomposing the velocity gradient tensor $L_{ij}$ according to

$$L_{ij} = \nabla_j \dot{X}_i = \Gamma_{ij} + \Xi_{ij} \qquad 8$$

where $\Gamma_{ij} = \frac{1}{2}[L_{ij}+L_{ji}]$, $\Xi_{ij} = \frac{1}{2}[L_{ij}-L_{ji}]$. The velocity gradient $L_{ij}$ is given with respect to the particle's local coordinate axis and is thus a function of the independent particle orientation angle vector $\underline{\Theta} = [\phi \quad \theta \quad \psi]^T$ obtained by a transformation operation according to

$$L_{ij} = Z_{X_{mi}} Ł_{mn} Z_{X_{nj}} \qquad 9$$

where $Ł_{ij}$ is the velocity gradient in the global reference frame axis. The transformation tensor $Z_{\theta_{ij}}$ is given in terms of the Euler angles as:

$$Z_{X_{ij}} = \Pi_{mi}^{(1)} \Pi_{nm}^{(2)} \Pi_{jn}^{(3)} \qquad 10$$

where,

$$\Pi_{ij}^{(k)} = \delta_{in}\delta_{jn} + (1-\delta_{in})(1-\delta_{jn})[\delta_{ij}\cos\Theta_k + (j-i)\sin\Theta_k], \qquad n = 2 + -1^k \qquad 11$$

At the particle's surface, the field velocity is given by

$$\dot{X}_i^p = \dot{X}_i\big|_{\lambda=0} = \epsilon_{ijk}\Psi_j X_k \qquad 12$$

The particle's angular velocity $\Psi_i$ in the local reference frame is given by the expression.

$$\Psi_i = \Xi_i + M_{ij} D_j \qquad 13$$

where $\Xi_i$ is the vorticity vector, $D_i$ contains non-diagonal terms of the symmetric rate of deformation tensor $\Gamma_{ij}$, i.e.

$$D_k = \Gamma_{ij}, \qquad k = 6 - i - j \,|\, i \neq j$$

and the constant coefficient matrix $M_{ij}$ is defined as

$$M_{ij} = M_i \delta_{ij}, \qquad M_1 = \frac{b^2-c^2}{b^2+c^2}, \qquad M_2 = \frac{c^2-a^2}{c^2+a^2}, \qquad M_3 = \frac{a^2-b^2}{a^2+b^2} \qquad 14$$

The angular velocities in the global reference coordinate axis $\underline{\dot{\Theta}}$ based on Euler's definition are obtained by the transformation operation

$$Z_{\Theta_{ij}} \dot{\Theta}_j = \Psi_i \qquad 15$$

where the transformation operator $\underline{\underline{Z}}_\Theta$ is given as (cf. Figure 1a) for the Euler definition of orientation angles

$$\underline{\underline{Z}}_\Theta = \begin{bmatrix} \cos\theta & 0 & 1 \\ -\sin\theta\cos\psi & \sin\psi & 0 \\ \sin\theta\sin\psi & \cos\psi & 0 \end{bmatrix} \qquad 16$$

Figure 1a illustrates the ellipsoidal particle of interest suspended in simple shear flow as shown. The normal and shear stress components at any point in the flow field may be evaluated for incompressible fluid as

$$\sigma_{ij} = -p\delta_{ij} + \mu_1[\nabla_i \dot{X}_j + \nabla_j \dot{X}_i] \qquad 17$$

On the particle's surface, the stress reduces to $\sigma_{ij} = -p\delta_{ij}$ implying that the only active stresses on the particle's surface are the hydrostatic pressure acting normal to the surface.

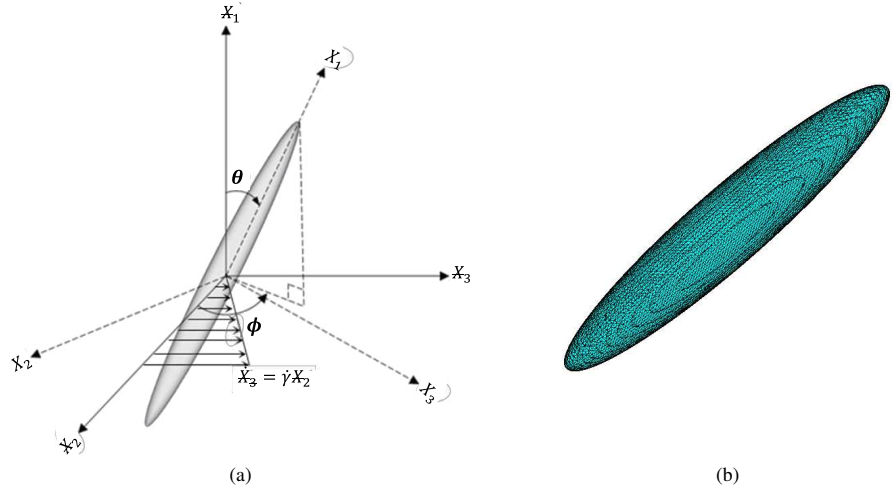

(a)       (b)
Figure 1: (a) Fiber orientation angles definition (b) Mesh refinement on the fiber surface.

Our main interest here is to evaluate the motion, and surface pressure and velocity of the ellipsoidal inclusion using Jeffery's equations given above. To compute surface pressure and velocity distribution on the particle surface, the ellipsoidal surface is discretized using MATLAB's inbuilt PDE modeller (MathWorks, Natick, MA, USA) where vertices were imposed at ends of the ellipsoid to enable the calculation of particle tip pressure (cf. Figure 1b). At the mesh points, the flow-field pressure and velocities are evaluated using eqn. 1 & 2 respectively. The degree of mesh refinement is critical to obtaining accurate pressure extremities and locations on the particle surface. A 4$^{th}$ order explicit Runge-Kutta ordinary differential equation (ODE) technique is used to numerically integrate the particle's angular velocities (cf. eqn. 13) with time to obtain solutions of the particle orientation angles, and the associated field state (pressure and velocities on each node of the particle surface) based on Jeffery's model equations.

**Homogenous Flow Considerations**

Various homogenous flows similar to those used in short fiber composite fiber orientation simulations[72] are considered here which serve as input for our particle motion studies. These homogenous flows also serve as a basis for understanding the flow fields development in common extrusion-deposition additive manufacturing (EDAM) polymer composite processing that involves a combination of shearing and extensional components within the flow (cf. Appendix B). The following flows are considered in this study:

(i)      Simple Shear flow (SS), i.e., $L_{23} = \dot{\gamma}$
(ii)     Two Stretching/Shearing flows (SUA), including simple shear in $X_2X_3$ plane superimposed with uniaxial elongation in the $X_3$-direction, i.e., $L_{11} = L_{22} = -\dot{\varepsilon}$, $L_{33} = 2\dot{\varepsilon}$, $L_{23} = \dot{\gamma}$. Two cases are considered, balanced shear/stretch, $\dot{\gamma}/\dot{\varepsilon} = 10$, and dominant stretch, $\dot{\gamma}/\dot{\varepsilon} = 1$
(iii)    Uniaxial Elongation flow (UA) in the $X_3$ direction, i.e., $L_{11} = L_{22} = -\dot{\varepsilon}$, $L_{33} = 2\dot{\varepsilon}$
(iv)    Biaxial Elongation (BA) flow in the $X_2 - X_3$ plane, i.e., $L_{11} = -2\dot{\varepsilon}$, $L_{22} = L_{33} = \dot{\varepsilon}$

(v) Two shear/planar-elongation flows (PST), including simple shear in $X_2 - X_3$ plane superimposed on planar elongation in $X_1 - X_3$ plane, i.e., $L_{11} = -\dot{\varepsilon}, L_{33} = \dot{\varepsilon}, L_{23} = \dot{\gamma}$. Two cases are considered including balanced shear-planar elongation with $\dot{\gamma}/\dot{\varepsilon} = 10$, and dominant planar elongation with $\dot{\gamma}/\dot{\varepsilon} = 1$.

(vi) Balanced shear/bi-axial elongation flow (SBA), simple shear in the $X_2 - X_3$ plane superimposed on biaxial elongation, i.e., $L_{33} = \dot{\varepsilon}$, $L_{22} = \dot{\varepsilon}, L_{23} = \dot{\gamma}, L_{11} = -2\dot{\varepsilon}$. Two cases are considered which include $\dot{\gamma}/\dot{\varepsilon} = 1$ and $\dot{\gamma}/\dot{\varepsilon} = 10$

(vii) Triaxial Elongation flow (TA), i.e., $L_{11} = L_{22} = L_{33} = \dot{\varepsilon}$

(viii) Balanced shear/tri-axial elongation flow (STA), including simple shear in the $X_2 - X_3$ plane superimposed on biaxial elongation, i.e., $L_{11} = L_{22} = L_{33} = \dot{\varepsilon}, L_{23} = \dot{\gamma}$, Two cases are considered i.e. $\dot{\gamma}/\dot{\varepsilon} = 1$, and $\dot{\gamma}/\dot{\varepsilon} = 10$

Classification of the various combined homogenous flow regimes based on the flow parameter $\bar{\nu}$ (cf. Appendix B) is given in Table I below

| $\dot{\gamma}/\dot{\varepsilon}$ | SUA | PST | SBA | STA |
|---|---|---|---|---|
| 1 | 0.5657 | 0.3820 | 0.5657 | 0.4514 |
| 10 | 0.0283 | 0.0098 | 0.0283 | 0.0146 |

Table I: Flow parameter values $\bar{\nu}$ for the combined homogenous flow types

For visualization purposes and to better interpret the results that follows in later section, typical flow streamlines around a particle suspended in the mixed mode flow conditions are presented in Figure 2. In all flow types, simple shear is applied in the $X_2 - X_3$ plane and the particle is initially oriented in the $X_2$ direction. The SUA flow (cf. Figure 2a) tends to orient the particle such that its major axis aligns with the $X_3$ direction of stretching, thus mitigating the tumbling motion in the $X_2 - X_3$ shear plane that occurs under simple shear flow alone. The inward flow in the y-direction initially accelerates the particle, aiding the tumbling motion into the direction of applied extension. High shear to extension rate dominance is thus required to prevent the particle from stalling in the $X_3$ direction. In the PST flow type shown in Figure 2b, the $X_1$ direction inward flow tends to constrain particle tumbling motion in the $X_2 - X_3$ shear plane and promotes preferential alignment of the particle in the z-direction and there is no flow in the y-direction that influence the particles initial motion. Unlike the SUA flow condition, in the SBA flow regime (cf. Figure 2c), the $X_1$ direction inward flow limits particle tumbling motion in the $X_2 - X_3$ shear plane without promoting directional preference for the particle alignment in the shear plane. Hence there is no tendency for particle stall to occur irrespective of the shear-extension rate dominance. Since the STA flow type has equal applied extension in all principal directions, the deviator of the velocity gradient has no principal component, and the particle's behavior under this flow type is similar to that under simple shear flow.

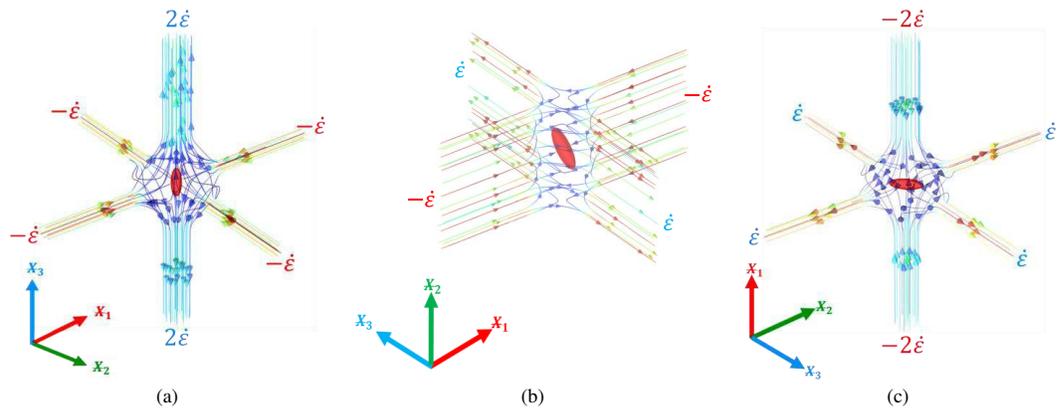

(a)      (b)      (c)

Figure 2: Visualization of the suspended particle in the combined shearing in $X_2 - X_3$ plane and (a) uniaxial elongation (SUA), (b) planar stretching (PST), and (c) biaxial elongation (SBA) flow conditions.

For the case of an axisymmetric ellipsoidal particle suspended in unconfined simple shear flow (see type (i) flow above) with velocity gradient $L_{23} = \dot{\gamma}$, Jeffery[15] derived analytical expressions for the particle's angular velocities given as

$$\dot{\phi}(t) = \frac{\dot{\gamma}}{2}[\kappa \cos 2\phi + 1], \qquad \dot{\theta}(t) = \frac{\dot{\gamma}}{2}\frac{(\kappa \sin 2\phi)\sqrt{(\kappa \cos 2\phi + 1)\varsigma^2(1+\kappa)}}{[(\kappa \cos 2\phi + 1) + \varsigma^2(1+\kappa)]}, \qquad \dot{\psi}(t) = -\frac{\dot{\gamma}}{2}(\kappa \cos 2\phi) \cos \theta \qquad 18$$

where the precession $\dot{\phi}$ is observed to be independent of $\theta$ and $\varsigma$ is the orbit constant. By integrating the angular velocities, Jeffery further obtained expressions for the corresponding particle orientation angles which may be written as

$$\phi(t) = \tan^{-1}\left\{\sqrt{\frac{1+\kappa}{1-\kappa}}\tan\left[\frac{\dot{\gamma}}{2}\sqrt{1-\kappa^2}\,t\right]\right\}, \qquad \theta(t) = \tan^{-1}\left\{\varsigma\frac{\sqrt{1+\kappa}}{\sqrt{\kappa \cos 2\phi + 1}}\right\}, \qquad \psi(t) = \int_0^t \left(\frac{\dot{\gamma}}{2} - \dot{\phi}\right)\cos\theta\, dt \qquad 19$$

where $\dot{\gamma}$ is the shear-rate, $\kappa$ is a shape factor given as $\kappa = (r_e^2 - 1)/(r_e^2 + 1)$. The orbit constant of integration $\varsigma$ can be shown to become $\varsigma = \tan \theta_0$ when $\phi_0 = 0$ and $\theta_0 \leq \theta \leq \tan^{-1}\{r_e \varsigma\}$[15]. For in-plane particle rotation, $\varsigma = +\infty$ such that $\theta = \pi/2$, $\psi = 0, \dot{\psi} = \dot{\theta} = 0$. The corresponding period for the in-plane particle tumbling motion in simple shear flow about the ellipsoid's polar axis is

$$\tau_1 = \frac{4\pi}{\dot{\gamma}\sqrt{1-\kappa^2}} \qquad 20$$

As the ellipsoid rotates in the $X_2 - X_3$ plane of shear flow, $\dot{\phi}$ reaches a maximum value when the particle is oriented normal to the principal direction of the fluid motion, i.e., at $\phi = n\pi$, $|n| \geq 0$ (cf. Figure 1a), and attains a minimum value when it aligns in the flow direction i.e., at $\phi = n\pi/2, |n| \geq 1$[76]. The limit of the precession is thus $0 \leq \dot{\phi} \leq \dot{\gamma}$ for ellipsoidal particles and $\dot{\phi} = \dot{\gamma}/2$ for spherical particles. The extremum of the nutation $\dot{\theta}$ occurs when $\phi = Re\{.5\cos^{-1}q\}$, where $q$ is the solution to the cubic equation defined as

$$\{q : \kappa^2 q^3 + 3\kappa(B+1)q^2 + (\kappa^2 + 2B + 2)q + \kappa(1-B) = 0\}, \qquad B = \varsigma^2(1+\kappa) \qquad 21$$

The nutation ranges between $-\dot{\gamma}/4 \leq \dot{\theta} \leq \dot{\gamma}/4$ for spheroidal particles, and it is critical for rodlike particles when $\varsigma = 1/\sqrt{2}$, and for disc-like particles when $\varsigma = +\infty$. It attains a value of $\dot{\theta} = 0$ for spherical particles. Likewise, the particle spin rate, $\dot{\psi}$ reaches a minimum at $\phi = n\pi, n \geq 0$, and a maximum value at $\phi = .5\cos^{-1}\left\{\left[-(3B+4) \pm \sqrt{B(9B+8)}\right]/4\kappa\right\}$. The spin-rate ranges between $-\dot{\gamma}/2 \leq \dot{\psi} \leq \dot{\gamma}/2$ and it is critical for rod-like particles when $\varsigma = 0$ and for disc-shaped particle when $\varsigma = +\infty$.

We now consider a more complicated flow condition and derive expressions for the case of an axisymmetric particle suspended in combined elongation and shear flow, i.e., flow types (ii, v, vi, & viii) given above following similar procedures adopted by Jeffery[15] for the case of simple shear flow. Consider a flow with velocity gradient of the form

$$\underline{\underline{L}} = \begin{bmatrix} \dot{\varepsilon}_1 & 0 & 0 \\ 0 & \dot{\varepsilon}_2 & 0 \\ 0 & \dot{\gamma} & \dot{\varepsilon}_3 \end{bmatrix} \qquad 22$$

where the $trace\left(\underline{\underline{L}}\right) = 0$, i.e., $\dot{\varepsilon}_1 + \dot{\varepsilon}_2 + \dot{\varepsilon}_3 = 0$. It can be shown that the angular velocities of a particle for this $\underline{\underline{L}}$ may be written as

$$\begin{bmatrix} \dot{\phi} \\ \dot{\theta} \\ \dot{\psi} \end{bmatrix} = \begin{bmatrix} \frac{\dot{\gamma}}{2} + \frac{\kappa}{2}\{\dot{\gamma}\cos 2\phi - [\dot{\varepsilon}_2 - \dot{\varepsilon}_3]\sin 2\phi\} \\ \frac{\kappa}{4}\{\dot{\gamma}\sin 2\phi + [\dot{\varepsilon}_2 - \dot{\varepsilon}_3]\cos 2\phi - [2\dot{\varepsilon}_1 - \dot{\varepsilon}_2 - \dot{\varepsilon}_3]\}\sin 2\theta \\ -\frac{\kappa}{2}\{\dot{\gamma}\cos 2\phi - [\dot{\varepsilon}_2 - \dot{\varepsilon}_3]\sin 2\phi\}\cos\theta \end{bmatrix} \qquad 23$$

where the in-plane angular velocity reduces to

$$\dot{\phi} = \frac{d\phi}{dt} = \frac{1}{2}\{\dot{\gamma}(1 + \kappa\cos 2\phi) - [\dot{\varepsilon}_2 - \dot{\varepsilon}_3]\kappa\sin 2\phi\} \qquad 24$$

By integrating $\dot{\phi}$ in eqn. 24, we obtain an expression for the in-plane orientation angle $\phi$ in these flow-types with characteristics velocity gradient $\underline{\underline{L}}$ given as

$$\tan\phi = \frac{k\kappa}{\kappa-1}\tan\left[\tan^{-1}\frac{1}{k}\left[\frac{\dot{\varepsilon}_2-\dot{\varepsilon}_3}{\dot{\gamma}} + \frac{\kappa-1}{\kappa}\tan\phi_0\right] - \frac{1}{2}k\kappa\dot{\gamma}t\right] - \frac{\kappa}{\kappa-1}\frac{\dot{\varepsilon}_2-\dot{\varepsilon}_3}{\dot{\gamma}} \qquad 25$$

where,

$$k = \sqrt{\frac{1}{\kappa^2} - \frac{\overline{\dot{\varepsilon}_2-\dot{\varepsilon}_3}^2}{\dot{\gamma}} - 1} \qquad 26$$

If the initial orientation $\phi_0 = 0$, then eqn. 25 reduces to

$$\tan\phi = -\frac{\kappa}{\kappa-1}\left[\frac{k^2 + \left[\frac{\dot{\varepsilon}_2-\dot{\varepsilon}_3}{\dot{\gamma}}\right]^2}{k\cot[.5k\kappa\dot{\gamma}t] + \frac{\dot{\varepsilon}_2-\dot{\varepsilon}_3}{\dot{\gamma}}}\right] \qquad 27$$

By integrating $\dot{\theta}$ in eqn. 23, we can directly obtain an expression for $\theta$ as

$$\tan\theta = \left[\frac{\frac{1}{\kappa}+\cos 2\phi_0 - \frac{[\dot{\varepsilon}_2-\dot{\varepsilon}_3]}{\dot{\gamma}}\sin 2\phi_0}{\frac{1}{\kappa}+\cos 2\phi - \frac{[\dot{\varepsilon}_2-\dot{\varepsilon}_3]}{\dot{\gamma}}\sin 2\phi}\right]^{1/2} \tan\theta_0\, e^{-\kappa/2[2\dot{\varepsilon}_1-\dot{\varepsilon}_2-\dot{\varepsilon}_3]t} \qquad 28$$

It can be shown that for the special case of initial polar orientation angle $\phi_0 = 0$, then eqn. 28 reduces to

$$\tan\theta = \left[\frac{\frac{1}{\kappa}+1}{\frac{1}{\kappa}+\cos 2\phi - \frac{[\dot{\varepsilon}_2-\dot{\varepsilon}_3]}{\dot{\gamma}}\sin 2\phi}\right]^{1/2} \tan\theta_0\, e^{-\kappa/2[2\dot{\varepsilon}_1-\dot{\varepsilon}_2-\dot{\varepsilon}_3]t} \qquad 29$$

Further, the spin $\psi(t)$ for these flow conditions may be written in integral form as

$$\psi(t) = \int_0^t \left(\frac{\dot{\gamma}}{2}-\dot{\phi}\right)\cos\theta\, dt \qquad 30$$

The quarter-period of rotation may be derived from eqn. 27 by finding the pole of the above expression of $\tan\phi$ as

$$\tau_1^{0.25} = \frac{2}{k\kappa\dot{\gamma}}\left[\pi - \tan^{-1}\left[\frac{k\dot{\gamma}}{\dot{\varepsilon}_2-\dot{\varepsilon}_3}\right]\right] \qquad 31$$

The period for a complete tumbling motion in this flow type is obtained by finding the zero of $\tan\phi$ in eqn. 27 above which is given as

$$\tau_1 = \frac{4\pi}{k\kappa\dot{\gamma}} \qquad 32$$

When $(\dot{\varepsilon}_2-\dot{\varepsilon}_3)/\dot{\gamma} = 0$, the flow is essentially simple shear, and the period is as given in eqn. 20 above. The particle stalls when $k^2 \leq 0$, i.e., when

$$\frac{\dot{\varepsilon}_2-\dot{\varepsilon}_3}{\dot{\gamma}} \geq \frac{\sqrt{1-\kappa^2}}{\kappa} \qquad 33$$

and the stall angle $\phi_s$ is derived by equating $\dot{\phi} = 0$ (cf. eqn. 24) to obtain

$$\tan 2\phi_s = \left[\frac{\dot{\varepsilon}_2-\dot{\varepsilon}_3}{\dot{\gamma}} \pm i\frac{k}{\kappa}\right]\bigg/\left[\frac{\overline{\dot{\varepsilon}_2-\dot{\varepsilon}_3}^2}{\dot{\gamma}} - \frac{1}{\kappa^2}\right], \qquad \phi_s = \begin{cases}\phi_s + \pi/2, & \phi_s < 0 \\ \phi_s, & \phi_s \geq 0\end{cases} \qquad 34$$

Correspondingly, given a stall angle tolerance $\phi_{tol}$, the time for particle stall is obtained by equating eqn. 27 and 34, i.e. $t_s : \phi(t_s) = \phi_s - \phi_{tol}$. When eqn. 34 is satisfied ($k = 0$), the stall angle may be shown to be

$$\phi_{onset} = \tan^{-1} r_e \qquad 35$$

The particle orientation at stall for the special class of homogenous flows (described as ii, v, vi, and viii above) can be obtained by using Newton-Raphson numerical iterative process to zero the angular velocities thus

$$\underline{\Theta}_s^{\rho+} = \underline{\Theta}_s^{\rho-} - \underline{\underline{J}}_{\Theta_1}^{-1}\underline{\dot{\Theta}}^\rho \qquad 36$$

where $\underline{\Theta}_s^{\rho-} = [\phi_s \quad \theta_s]^T$, $\underline{\dot{\Theta}}^\rho = [\dot{\phi} \quad \dot{\theta}]^T$, and the Jacobian $\underline{\underline{J}}_\Theta$ is given as

$$\underline{\underline{J}}_{\Theta_1} = \begin{bmatrix} -4\dot{\theta}\csc 2\theta - \kappa[2\dot{\varepsilon}_1 - \dot{\varepsilon}_2 - \dot{\varepsilon}_3] & 0 \\ \left\{\dot{\phi} - \dfrac{\dot{\gamma}}{2}\right\}\sin 2\theta & 2\dot{\theta}\cot 2\theta \end{bmatrix} \qquad 37$$

For particle motion in more general class of Newtonian homogenous flows with velocity gradient $\underline{\underline{L}}$ the stall angle can be obtained using the Newton-Raphson procedure in Appendix C.

Jeffery's model derivations are limited to the standard assumption of single rigid ellipsoidal shaped particle suspended in Newtonian viscous linear homogenous flows. Practically speaking, the pressure driven flow of polymer melt through EDAM nozzle contraction during material processing is more accurately characterized by a quadratic ambient flow-field such as given in Lubansky et al [75]. As such, development of a more realistic solution would involve a velocity gradient with higher order polynomial terms which is a relevant direction for future studies. For more general conditions, it is common to employ the Finite Element Analysis (FEA) which are not bound by the limitations of the Jeffery's model and can include inter and intra fibre forces, non-ellipsoidal fibre shape, non-Newtonian visco-elastic fluid rheology, confinement flows, and other deviations from standard conditions. Moving beyond Jeffery's model assumptions may result in a preferred particle configuration that is independent of its initial orientation and may cause the particle to align with the flow or vorticity direction[19-21]. In the sections following we describe an FEA modelling approach that may be used to investigate the effect of Generalized Newtonian Fluid (GNF) rheology on the particle dynamics and surface pressure response.

**FEA Single Particle Model with GNF Rheology**

In the FEA model analysis present here, we simulate the motion of a single rigid spheroidal particle suspended in homogenous viscous flow with GNF rheology. The flow domain $\vartheta$ for the single particle micromodel analysis is shown in Figure 3a. The model extends the Newtonian fluid single fiber model developed by Zhang et. al.[69,68,70] and implemented by Awenlimobor et al.,[63,64] to simulate GNF flow. In this approach, the governing equations are based on the Stokes assumption of creeping, incompressible, isothermal, steady state, low Reynolds number viscous flow where the mass and momentum conservation equations may be written as

$$\nabla_i \dot{X}_i = 0 \qquad 38$$
$$\nabla_i \sigma_{ij} + f_j = 0 \qquad 39$$

In the above, $\nabla_i$ is the gradient operator, $\dot{X}_i$ is the flow velocity vector, $f_j$ is the body force vector, and $\sigma_{ij}$ is the Cauchy stress tensor given as

$$\sigma_{ij} = \tau_{ij} - p\delta_{ij} \qquad 40$$

In eqn. 40, $p$ is the hydrostatic fluid pressure, $\delta_{ij}$ is the kronecker delta, and $\tau_{ij}$ is the deviatoric stress tensor defined in terms of the strain rate tensor $\dot{\gamma}_{ij}$ by the constitutive relation

$$\tau_{ij} = 2\mu(\dot{\gamma})\dot{\gamma}_{ij} \qquad 41$$

where the viscosity $\mu$ is considered to be a function of the strain rate magnitude $\dot{\gamma} = \sqrt{2\dot{\gamma}_{ij}\dot{\gamma}_{ji}}$. The simulations presented below solve eqns. 38-41 for quasi-steady velocity and pressure within the fluid domain surrounding the ellipsoidal inclusion using our custom finite element analysis (FEA) program developed in MATLAB. We assume a non-porous particle surface with zero slip allowance and velocity boundary conditions are prescribed with respect to the particle's local coordinate reference axes.

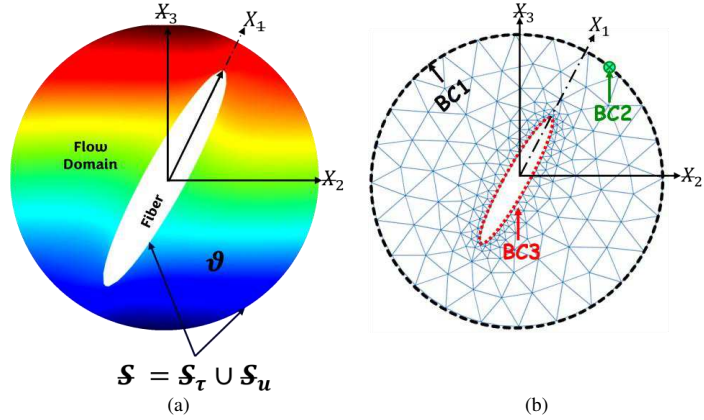

(a)        (b)

Figure 3: FEA model showing (a) flow domain (b) prescribed boundary conditions

Similar to previous single particle Newtonian fluid analyses[63], the velocities and velocity gradients of the prevailing flow are used to compute the far-field velocities on the fluid domain boundary $\dot{X}_i^{BC1}$ (cf. Figure 3b) of the micromodel as

$$\dot{X}_i^{BC1} = \dot{X}_i^\infty = Z_{X_{ji}}\dot{X}_j^\psi + Z_{X_{mi}}L_{mn} Z_{X_{nj}}\Delta X_j^{BC1} \qquad 42$$

where $Z_{X_{ji}}$ is the local to global transformation tensor, $\dot{X}_j^\psi$ is the flow-field velocity vector, $L_{mn}$ is the velocity gradient tensor in global reference frame and $\Delta X_j$ is the position vector with respect to the particle's center. Again, referring to Figure 3b, the velocity on the particle's surface $\dot{X}_i^{BC3}$ is computed from the particle's center translational and rotational velocities assuming rigid body motion which is written with respect to the particle's local reference axis as

$$\dot{X}_i^{BC3} = \dot{X}_i^p = Z_{X_{ji}}\dot{X}_j^c + \epsilon_{ijk}Z_{\Theta_{jn}}\dot{\Theta}_n \Delta X_k^{BC3} \qquad 43$$

where $\dot{X}_i^c$ is the particle's center translational velocity vector and $\dot{\Theta}_i$ is the particle's angular velocity vector. A pressure point constraint $p_{BC2}$ is imposed at a node on the far-field fluid domain (see, e.g., BC2 in Figure 3b) with a magnitude equal to the prescribed static fluid pressure $p_0$, i.e.

$$p_{BC2} = p_0 \qquad 44$$

We define a fluid domain size factor $\mathbb{M} = d_f/2c$[63] (where $d_f$ is the diameter of the flow domain and $c$ is the major axis length of the particle). The flow domain size thus increases linearly with the size of the particle. In our analysis, we utilize a factor of $\mathbb{M} = 10$ which is determined to be sufficiently large to yield accurate results. The fluid domain discretization for the base case having a particle geometric aspect ratio $r_e = 6$ appears in Figure 4a & b where an increasing mesh density is used near the particle and particles tip. All FEA simulations are performed with a 10-node quadratic, iso-parametric tetrahedral serendipity element as shown in Figure 4c.

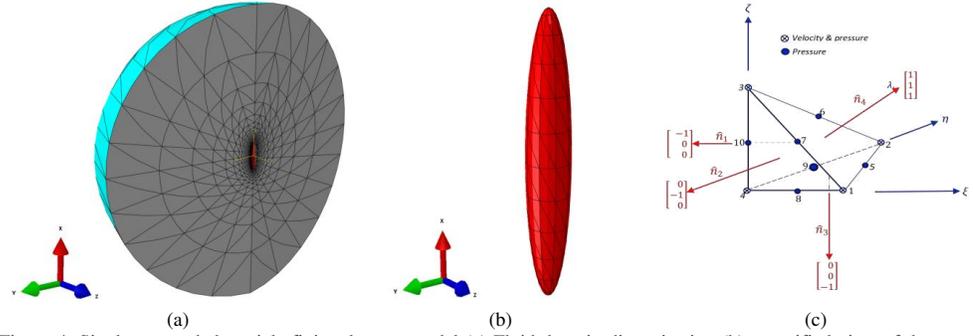

(a)      (b)      (c)

Figure 4: Single suspended particle finite element model (a) Fluid domain discretization (b) magnified view of the domain mesh on the surface of the rigid particle (c) element selection with active degrees of freedom.

The weak form of the nonlinear finite element equations may be transformed in the usual manner to a system of algebraic equations written in terms of the solution variable vector $\underline{u}$ and the global system residual vector $\underline{\Sigma}$ as

$$\underline{\Sigma} = \underline{\underline{K}}(\underline{u})\underline{u} - \underline{f} \qquad 45$$

where $\underline{\underline{K}}$ is the global system 'stiffness' matrix, $\underline{u} = [\underline{v} \quad \underline{p}]^T$ is the primary solution vector containing nodal velocities $\underline{v}$ and pressures $\underline{p}$ and $\underline{f}$ is the secondary variable vector containing the associated nodal reaction forces and flow rates. To simplify the solution procedure, the global system matrix is partitioned into essential $'e'$ (known) and free $'f'$ (unknown) degrees of freedom (dofs) as

$$\underline{\Sigma} = \begin{Bmatrix} \underline{\Sigma}_f \\ \underline{\Sigma}_e \end{Bmatrix} = \begin{Bmatrix} \underline{\underline{K}}_{ff} & \underline{\underline{K}}_{fe} \\ \underline{\underline{K}}_{ef} & \underline{\underline{K}}_{ee} \end{Bmatrix} \begin{Bmatrix} \underline{u}_f \\ \underline{u}_e \end{Bmatrix} - \begin{Bmatrix} \underline{f}_f \\ \underline{f}_e + \underline{g}_e \end{Bmatrix} \qquad 46$$

where $\underline{u}_f$ & $\underline{g}_e$ are the unknowns quantities to be computed in the finite element analysis. The unknown free velocity and pressure dofs in $\underline{u}_f$ are computed via a Newton Raphson iterative algorithm by zeroing the free residual vector $\underline{\Sigma}_f$. i.e $\underline{u}_f$ is iteratively updated until it approaches the actual solution according to

$$\underline{u}_f^+ = \underline{u}_f^- - \underline{\underline{J}}_{ff}^{-1}\underline{\Sigma}_f \qquad 47$$

In the above, the Tangent Stiffness Matrix (TSM) or Jacobian $\underline{\underline{J}}_{ff}$ is obtained by differentiating the free residual vector $\underline{\Sigma}_f$ defined in eqn. 46 with respect to the free degrees of freedom $\underline{u}_f$ to obtain

$$\underline{\underline{J}}_{ff} = \frac{\partial \underline{\Sigma}_f}{\partial \underline{u}_f} = \frac{\partial \underline{\underline{K}}_{ff}}{\partial \underline{u}_f}\underline{u}_f + \frac{\partial \underline{\underline{K}}_{fe}}{\partial \underline{u}_f}\underline{u}_e + \underline{\underline{K}}_{ff} - \frac{\partial \underline{f}_f}{\partial \underline{u}_f} \qquad 48$$

The unknown reactions forces and flow rates at the essential dofs in $\underline{g}_e$ are computed by setting the essential residual vector $\underline{\Sigma}_e = 0$ (cf. eqn. 46) to obtain as

$$\underline{g}_e = \underline{\underline{K}}_{ef}\underline{u}_f + \underline{\underline{K}}_{ee}\underline{u}_e - \underline{f}_e \qquad 49$$

The global residual vector and Jacobian are assembled from individual element residual $\underline{\Sigma}^e$ and element tangent stiffness matrices $\underline{\underline{J}}^e$ in the usual manner. The element residual vector $\underline{\Sigma}^e$ is written in terms of the FEA integral equations as

$$\underline{\Sigma}^e = \begin{Bmatrix} \underline{\Sigma}_1^e \\ \underline{\Sigma}_2^e \end{Bmatrix} = \left\{ \begin{array}{c} \int_{\vartheta^e} \underline{\omega}_1 (\underline{\nabla} \cdot \underline{v}) d\vartheta \\ \int_{\vartheta^e} (\underline{\underline{\nabla}}_s \cdot \underline{\omega}_2)^T \mu(\dot{\gamma})\underline{\underline{C}}_o (\underline{\underline{\nabla}}_s \cdot \underline{v}) d\vartheta - \int_{\vartheta^e} p(\underline{\nabla} \cdot \underline{\omega}_2) d\vartheta - \int_{\vartheta^e} \rho \underline{\omega}_2^T \underline{f} d\vartheta - \int_{S_T^e} \underline{\omega}_2^T \underline{t} \, dS \end{array} \right\} \qquad 50$$

where $\underline{\Sigma}_1^e$ & $\underline{\Sigma}_2^e$ are element residual vectors derived from mass and momentum conservation, respectively, $\underline{\omega}_1$ and $\underline{\omega}_2$ are the arbitrary FEA weighting functions on the continuity and momentum equation, respectively, $\underline{\nabla}$ and $\underline{\underline{\nabla}}_s$ are the gradient vector and symmetric gradient matrix operator, respectively, defined in[77], $\underline{p}$ and $\underline{v}$ are the pressure and velocity field variables, $\rho$ is the fluid density, $\mu(\dot{\gamma})$ is the non-Newtonian fluid viscosity, $\underline{\underline{C}}_\rho$ is a constant coefficient matrix, $\underline{\bar{t}}$ and $\underline{f}$ are the surface traction and the body force vectors, and $S_T^e$ and $\vartheta^e$ are the element surface and interior domains of integration, respectively.

The element TSM $\underline{\underline{J}}^e$ is obtained by differentiating the element residual vector $\underline{\Sigma}^e$ with respect to the element solution variables $\underline{u}^e$ which contains $\underline{p}^e$ and $\underline{v}^e$, i.e., $\underline{u}^e = \begin{bmatrix} \underline{v}^e & \underline{p}^e \end{bmatrix}^T$ and

$$\underline{\underline{J}}^e = \frac{\partial \underline{\Sigma}^e}{\partial \underline{u}^e} = \frac{\partial}{\partial \underline{u}^e}\begin{Bmatrix} \underline{\Sigma}_1^e \\ \underline{\Sigma}_2^e \end{Bmatrix}, \qquad \underline{\underline{J}}^{eT} = \begin{bmatrix} \frac{\partial}{\partial \underline{v}^e} & \frac{\partial}{\partial \underline{p}^e} \end{bmatrix}^T \begin{Bmatrix} \underline{\Sigma}_1^e \\ \underline{\Sigma}_2^e \end{Bmatrix}^T \qquad 51$$

First order Façade derivatives are used to approximate the tangent stiffness matrix according to

$$\frac{\partial \underline{\Sigma}}{\partial \underline{u}} \Delta \underline{u} = \underline{\Sigma}(\underline{u} + \Delta \underline{u}) - \underline{\Sigma}(\underline{u}), \qquad \underline{\Sigma} = \underline{\Sigma}(\underline{u}) \qquad 52$$

which we apply to the continuity residual term $\underline{\Sigma}_1^e$ to obtain derivatives with respect to the velocity and pressure as

$$\frac{d\underline{\Sigma}_1^e}{d\underline{v}} \Delta \underline{v} = \int_{\vartheta^e} \underline{\omega}_1 (\underline{\nabla} \cdot \Delta \underline{v}) \, d\vartheta, \qquad \frac{d\underline{\Sigma}_1^e}{d\underline{p}} \Delta \underline{p} = 0 \qquad 53$$

Similarly, derivatives of the momentum conservation term with respect to the solution variables after algebraic manipulations are, respectively, given as

$$\frac{d\underline{\Sigma}_2^e}{d\underline{v}} \Delta \underline{v} = \int_{\vartheta^e} (\underline{\nabla}_s \underline{\omega}_2)^T \mu \underline{\underline{C}}_\rho \underline{\nabla}_s \Delta \underline{v} \, d\vartheta + \int_{\vartheta^e} \frac{1}{\mu^2} \frac{1}{\dot{\gamma}} \frac{\partial \mu}{\partial \dot{\gamma}} \Big[ (\underline{\nabla}_s \underline{\omega}_2)^T \mu \underline{\underline{C}}_\rho \underline{\nabla}_s \underline{v} \Big] \Big[ (\underline{\nabla}_s \underline{v})^T \mu \underline{\underline{C}}_\rho \underline{\nabla}_s \Delta \underline{v} \Big] d\vartheta \qquad 54$$

$$\frac{d\underline{\Sigma}_2^e}{d\underline{p}} \Delta \underline{p} = -\int_{\vartheta^e} (\underline{\nabla} \cdot \underline{\omega}_2) \Delta \underline{p} \, d\vartheta \qquad 55$$

It follows that the Galerkin formulation written as the element residual vector $\underline{\Sigma}^e$ and tangent stiffness matrix $\underline{\underline{J}}^e$ in tensorial representation are given respectively as

$$\underline{\Sigma}^e = \begin{Bmatrix} \int_{\vartheta^e} \underline{\underline{B}}_s^{eT} \mu(\dot{\gamma}) \underline{\underline{C}}_\rho \underline{\underline{B}}_s^e d\vartheta & -\int_{\vartheta^e} \underline{B}^{eT} \underline{\Phi}^e d\vartheta \\ -\int_{\vartheta^e} \underline{\Phi}^{eT} \underline{B}^e d\vartheta & \underline{\underline{0}} \end{Bmatrix} \begin{Bmatrix} \underline{v}^e \\ \underline{p}^e \end{Bmatrix} - \begin{Bmatrix} \int_{\vartheta^e} \rho \underline{\underline{N}}^{eT} \underline{f} \, d\vartheta + \int_{S_T^e} \underline{N}^{eT} \underline{\bar{t}} \, dS \\ \underline{0} \end{Bmatrix} \qquad 56$$

and

$$\underline{\underline{J}}^e = \frac{d\underline{\Sigma}^e}{d\underline{u}^e} = \begin{Bmatrix} \int_{\vartheta^e} \underline{\underline{B}}_s^{eT} \mu \underline{\underline{C}}_\rho \underline{\underline{B}}_s^e d\vartheta + \int_{\vartheta^e} \frac{1}{\mu^2} \frac{1}{\dot{\gamma}} \frac{\partial \mu}{\partial \dot{\gamma}} (\underline{\underline{B}}_s^{eT} \mu \underline{\underline{C}}_\rho \underline{\underline{B}}_s^e \underline{v}^e) (\underline{v}^{eT} \underline{\underline{B}}_s^{eT} \mu(\dot{\gamma}) \underline{\underline{C}}_\rho^T \underline{\underline{B}}_s^e) d\vartheta & -\int_{\vartheta^e} \underline{B}^{eT} \underline{\Phi}^e d\vartheta \\ -\int_{\vartheta^e} \underline{\Phi}^{eT} \underline{B}^e d\vartheta & \underline{\underline{0}} \end{Bmatrix} \qquad 57$$

where

$\underline{\Phi}^e$ and $\underline{N}^e$ are the pressure and velocity interpolation functions, respectively,

$\underline{\underline{B}}^e$ and $\underline{\underline{B}}_s^e$ are 'strain' displacement matrices

$\underline{v}^e$ and $\underline{p}^e$ are respectively the velocities and pressures degrees-of-freedom (dof) at the respective element nodes

$S^e$ and $\vartheta^e$ are the element boundary surfaces and domain of integration, respectively.

In eqn. 57, $\dot{\gamma}$ is the scalar magnitude of the strain rate tensor $\underline{\underline{\dot{\gamma}}}$ which may be written in terms of FEA quantities as

$$\dot{\gamma} = \sqrt{\frac{1}{2} \underline{\underline{\dot{\gamma}}} : \underline{\underline{\dot{\gamma}}}} = \sqrt{(\underline{\nabla}_s \underline{v})^T \underline{\underline{C}}_\rho (\underline{\nabla}_s \underline{v})}, \qquad \dot{\gamma} = \sqrt{\underline{v}^{eT} \underline{\underline{B}}_s^{eT} \mu(\dot{\gamma}) \underline{\underline{C}}_\rho \underline{\underline{B}}_s^e \underline{v}^e} \qquad 58$$

In this work, we consider the non-Newtonian viscosity $\mu(\dot{\gamma})$ as that of a power-law shear-thinning fluid given as

$$\mu = m\dot{\gamma}^{n-1} \qquad 59$$

where $m$ is the flow consistency coefficient in $Pa \cdot s^n$ and $n$ is the power-law index, and $\dot{\gamma}$ is the scalar magnitude of the deformation tensor $\dot{\gamma}_{ij}$. In the second integral of the momentum equation Jacobian in eqn. 57 above, it is convenient to introduce a variable $\alpha = 1/(\mu^2 \dot{\gamma})\,(\partial\mu/\partial\dot{\gamma})$ to simplify the expression and make it generally applicable to other GNF fluids. It follows that $\alpha$ can be written for the power-law fluid as

$$\alpha = \frac{1}{\mu^2}\frac{1}{\dot{\gamma}}\frac{\partial\mu}{\partial\dot{\gamma}} = \frac{1}{\mu\dot{\gamma}^2}(n-1) \qquad 60$$

Alternatively, for a Carreau-Yasuda fluid, the expression for $\mu$ and $\alpha$ are, respectively,

$$\frac{\mu - \mu_\infty}{\mu_0 - \mu_\infty} = \{1 + (\lambda\dot{\gamma})^a\}^{(n-1)/a} \quad and \quad \alpha = \frac{1}{\mu^2}\frac{1}{\dot{\gamma}}\frac{\partial\mu}{\partial\dot{\gamma}} = \frac{1}{\dot{\gamma}^2}\frac{\mu - \mu_\infty}{\mu^2}\left\{\frac{n-1}{1 + (\lambda\dot{\gamma})^{-a}}\right\} \qquad 61$$

where, $\mu_0$ is the zero-shear viscosity, $\mu_\infty$ is an infinite-shear viscosity, $\lambda$ is a time constant, and a is a fitting parameter.

**Single Particle Motion with GNF Rheology**

In our numerical approach, the particle's motion is computed based on an appropriate explicit numerical ordinary differential equation solution technique by calculating its linear and rotational velocities that results in a zero net hydrodynamic force and torque acting on the particle's surface. Again, we adopt the Newton-Raphson's iterative method to determine the nonlinear solution of particle's translational and rotational velocities as

$$\underline{\dot{Y}}^+ = \underline{\dot{Y}}^- - \underline{\underline{J}}_H^{-1} \underline{\Sigma}_H \qquad 62$$

where $\underline{\dot{Y}}$ contains the particle's linear velocities $\underline{\dot{X}}^c$ and rotational velocity $\underline{\Psi}$, i.e., $\underline{\dot{Y}} = [\underline{\dot{X}}^c \quad \underline{\Psi}]^T$ and $\underline{\Sigma}_H$ is the particle hydrodynamic residual vector which is composed of the particle's hydrodynamic forces $\underline{F}_H$ and couple $\underline{Q}_H$, i.e., $\underline{\Sigma}_H = [\underline{F}_H \quad \underline{Q}_H]^T$ as a function of the particle's velocity, i.e., $\underline{\Sigma}_H = \underline{\Sigma}_H(\underline{\dot{Y}})$. Since calculations are performed with respect to the particle's local reference frame, the particle's velocity vector is transformed to global coordinate system according to the eqn. 63

$$\underline{\dot{\bar{Y}}} = \underline{\underline{Z}}_{\dot{Y}}\, \underline{\dot{Y}} \qquad 63$$

where variables on the global reference frame are accented by a strikethrough and the particle's velocity transformation tensor $\underline{\underline{Z}}_{\dot{Y}}$ is given by

$$\underline{\underline{Z}}_{\dot{Y}} = \begin{bmatrix} \underline{\underline{Z}}_X & \underline{0} \\ \underline{0}^T & \underline{\underline{Z}}_\Theta^{-1} \end{bmatrix} \qquad 64$$

We calculate the net hydrodynamic force vector $\underline{F}_H$ and couple $\underline{Q}_H$ on the particle by vector summation of the nodal reactions forces and torques on the particle surface as

$$\underline{F}_H = -\sum_k^{n_k} \underline{g}_e^{(k)}, \qquad \underline{Q}_H = -\sum_k^{n_k} \underline{\Delta X}^{(k)} \times \underline{g}_e^{(k)} \qquad 65$$

where $\underline{\Delta X}^{(k)}$, and $\underline{g}_e^{(k)}$ are the position vector and the nodal reaction force vector at the $k^{th}$ node on the particle surface ($BC3$), respectively, and $n_k$ is the total number of nodes on $BC3$. The particle hydrodynamic Jacobian $\underline{\underline{J}}_H$ in eqn. 62 above is obtained by differentiating the components of the particle hydrodynamic residual vector $\underline{\Sigma}_H$ with respect to components of the particle's velocity vector $\underline{\dot{Y}}$ as

$$\underline{\underline{J}}_H = \frac{\partial \underline{\Sigma}_H}{\partial \underline{\dot{Y}}} = \frac{\partial}{\partial \underline{\dot{Y}}}[\underline{F}_H \quad \underline{Q}_H]^T = \left[-\sum_k^{n_k}\frac{\partial \underline{g}_e^{(k)}}{\partial \underline{\dot{Y}}} \quad -\sum_k^{n_k}\underline{\Delta X}^{(k)}\times\frac{\partial \underline{g}_e^{(k)}}{\partial \underline{\dot{Y}}}\right]^T \qquad 66$$

Differentiating the global system FEA residual vector $\underline{\Sigma}$ in eqn. 46 with respect to the particle velocity vector $\underline{\dot{Y}}$ we obtain the derivative of the nodal reaction force vector $d\underline{g}_e/d\underline{\dot{Y}}$ in eqn. 66 as

$$\frac{d\underline{g}_e}{d\underline{\dot{Y}}} = \left\{\frac{\partial \underline{\underline{K}}_{ef}}{\partial \underline{u}_e}\underline{u}_f + \frac{\partial \underline{\underline{K}}_{ee}}{\partial \underline{u}_e}\underline{u}_e + 2\underline{\underline{K}}_{ee} - \frac{d\underline{f}_e}{d\underline{u}_e}\right\}\frac{d\underline{u}_e}{d\underline{\dot{Y}}} + \left\{\frac{\partial \underline{\underline{K}}_{ef}}{\partial \underline{\bar{u}}_f}\underline{u}_f + \frac{\partial \underline{\underline{K}}_{ee}}{\partial \underline{\bar{u}}_f}\underline{u}_e + 2\underline{\underline{K}}_{ef} - \frac{d\underline{f}_e}{d\underline{\bar{u}}_f}\right\}\frac{d\underline{u}_f}{d\underline{\dot{Y}}} \qquad 67$$

where the derivative $d\underline{u}_f/d\underline{\dot{Y}}$ is written in terms of the derivative $d\underline{u}_e/d\underline{\dot{Y}}$ as

$$\frac{d\underline{u}_f}{d\underline{Y}} = -\left\{\frac{\partial \underline{\underline{K}}_{ff}}{\partial \underline{\underline{u}}_f}\underline{u}_f + \frac{\partial \underline{\underline{K}}_{fe}}{\partial \underline{\underline{u}}_f}\underline{u}_e + 2\underline{\underline{K}}_{ff} - \frac{d\underline{f}_f}{d\underline{\underline{u}}_f}\right\}^{-1}\left\{\frac{\partial \underline{\underline{K}}_{ff}}{\partial \underline{u}_e}\underline{u}_f + \frac{\partial \underline{\underline{K}}_{fe}}{\partial \underline{u}_e}\underline{u}_e + 2\underline{\underline{K}}_{fe} - \frac{d\underline{f}_f}{d\underline{u}_e}\right\}\frac{d\underline{u}_e}{d\underline{Y}} \qquad 68$$

To obtain the FEA model derivatives in the above, we differentiate the global FEA system residual $\underline{\Sigma}$ in eqn. 47 with respect to the solution variable $\underline{u}$ to obtain the global FEA system Jacobian $\underline{\underline{J}}$ as

$$\underline{\underline{J}} = \frac{d\underline{\Sigma}}{d\underline{u}} = \begin{pmatrix}\underline{\underline{J}}_{ff} & \underline{\underline{J}}_{fe} \\ \underline{\underline{J}}_{ef} & \underline{\underline{J}}_{ee}\end{pmatrix} = \begin{cases}\left\{\frac{\partial \underline{\underline{K}}_{ff}}{\partial \underline{u}_f}\underline{u}_f + \frac{\partial \underline{\underline{K}}_{fe}}{\partial \underline{u}_f}\underline{u}_e + \underline{\underline{K}}_{ff} - \frac{\partial \underline{f}_f}{\partial \underline{u}_f}\right\} & \left\{\frac{\partial \underline{\underline{K}}_{ff}}{\partial \underline{u}_e}\underline{u}_f + \frac{\partial \underline{\underline{K}}_{fe}}{\partial \underline{u}_e}\underline{u}_e + \underline{\underline{K}}_{fe} - \frac{\partial \underline{f}_f}{\partial \underline{u}_e}\right\} \\ \left\{\frac{\partial \underline{\underline{K}}_{ef}}{\partial \underline{u}_f}\underline{u}_f + \frac{\partial \underline{\underline{K}}_{ee}}{\partial \underline{u}_f}\underline{u}_e + \underline{\underline{K}}_{ef} - \frac{\partial \underline{f}_e}{\partial \underline{u}_f}\right\} & \left\{\frac{\partial \underline{\underline{K}}_{ef}}{\partial \underline{u}_e}\underline{u}_f + \frac{\partial \underline{\underline{K}}_{ee}}{\partial \underline{u}_e}\underline{u}_e + \underline{\underline{K}}_{ee} - \frac{\partial \underline{f}_e}{\partial \underline{u}_e}\right\}\end{cases} \qquad 69$$

where eqn. 48 has been expanded to include all free and essential degrees of freedom in $\underline{u} = \{\underline{u}_f \quad \underline{u}_e\}^T$. In addition, the nodal reaction force vector derivative $d\underline{g}_e/d\underline{Y}$ in eqn. 67 is written in terms of the submatrices of the global FEA system Jacobian $\underline{\underline{J}}$ as

$$\frac{d\underline{g}_e}{d\underline{Y}} = \left\{\underline{\underline{K}}_{ee} + \underline{\underline{J}}_{ee}\right\}\frac{d\underline{u}_e}{d\underline{X}} + \left\{\underline{\underline{K}}_{ef} + \underline{\underline{J}}_{ef}\right\}\frac{d\underline{u}_f}{d\underline{Y}} \qquad 70$$

Likewise, the derivative $d\underline{u}_f/d\underline{Y}$ in eqn. 68 is also written in terms of the submatrices of the global system Jacobian $\underline{\underline{J}}$ as

$$\frac{d\underline{u}_f}{d\underline{Y}} = -\left\{\underline{\underline{K}}_{ff} + \underline{\underline{J}}_{ff}\right\}^{-1}\left\{\underline{\underline{K}}_{fe} + \underline{\underline{J}}_{fe}\right\}\frac{d\underline{u}_e}{d\underline{Y}} \qquad 71$$

Given the initial condition of the particle, $\underline{Y}^{j-1}$ at any instant with an associated velocity $\underline{\dot{Y}}^{j-1}$ at each $j^{th}$ time step, we update particle's position and orientation $\underline{Y}^j$ using on an explicit fourth order Runge-Kutta method. i.e.

$$\underline{Y}^j = \underline{Y}^{j-1} + \frac{\Delta t}{6}\left[\underline{\mathcal{K}}_1^{j-1} + 2\underline{\mathcal{K}}_2^{j-1} + 2\underline{\mathcal{K}}_3^{j-1} + \underline{\mathcal{K}}_4^{j-1}\right] \qquad 72$$

where

$$\begin{aligned}\underline{\mathcal{K}}_1^{j-1} &= f_Y(t^{j-1}, \underline{Y}^{j-1}) = \underline{\dot{Y}}^{j-1} & \underline{\mathcal{K}}_2^{j-1} &= f_Y(t^{j-1} + \Delta t/2, \underline{Y}^{j-1} + \Delta t/2\,\underline{\mathcal{K}}_1^{j-1}) \\ \underline{\mathcal{K}}_3^{j-1} &= f_Y(t^{j-1} + \Delta t/2, \underline{Y}^{j-1} + \Delta t/2\,\underline{\mathcal{K}}_2^{j-1}) & \underline{\mathcal{K}}_4^{j-1} &= f_Y(t^{j-1} + \Delta t, \underline{Y}^{j-1} + \Delta t\,\underline{\mathcal{K}}_3^{j-1})\end{aligned} \qquad 73$$

and the function $f_Y$ is used to evaluate the particles velocities $\underline{\dot{Y}}$ at time $t$ and position $\underline{Y}$

**Comparison of Jeffery's and FEA model**

To validate our FEA model-based particle motion simulations to calculations performed with Jeffery's equations, we first define the particle surface pressure $\bar{p}$ in dimensionless form as

$$\bar{p} = \frac{p - p_0}{\mu_1 \dot{\gamma}_c} \qquad 74$$

where $\dot{\gamma}_c$ is a characteristic strain rate of the flow-field. For a given $\mu_1$ and $\dot{\gamma}_c$, $\bar{p}$ is evaluated from eqn. 74 where $p$ is computed from Jeffery's model (cf. eqn. 1) and similarly from the nodal pressure solution of the FEA model described above. Likewise, the flow-field velocity magnitude is normalized with respect to the tangential velocity at the particle's tip is given as

$$\bar{v} = |\underline{\dot{x}}|/|\underline{\dot{X}}^t|, \qquad \underline{\dot{X}}^t = \underline{\theta} \times \underline{X}^t \qquad 75$$

where $\underline{X}^t$ is the position vector at particle's tip defined by the major axis length. To ensure consistency between the Jeffery's model equations and Finite Element Analysis (FEA) simulation results, we consider the particle's motion and surface pressure distribution for the case of a single rigid ellipsoidal particle suspended in viscous homogenous Newtonian (i.e., power-law index $n = 1$) flow. The FEA model is shown to exactly match Jeffery's results for a range of particle aspect ratios including $r_e = 1, 2, 3, 6,$ and $10$ (cf. Figure 5a for $\dot{\phi}$ and Figure 5b for $\bar{P}$). $\bar{P}$ is the dimensionless pressure at the particle's tip. Additionally, Jeffery's orbit exactly match our FEA results for the various flow conditions described above as shown in Figure 5c and Figure 5d which show components of the particle unit vector $\rho_i$, and maximum and minimum normalized surface pressure $\tilde{p}$. Results in Figure 5a & b are for one period of Jeffery's orbit, however, given that values at the end point exactly match within 0.25%, we expect the accuracy of our numerical approach to remain as particle rotations continue.

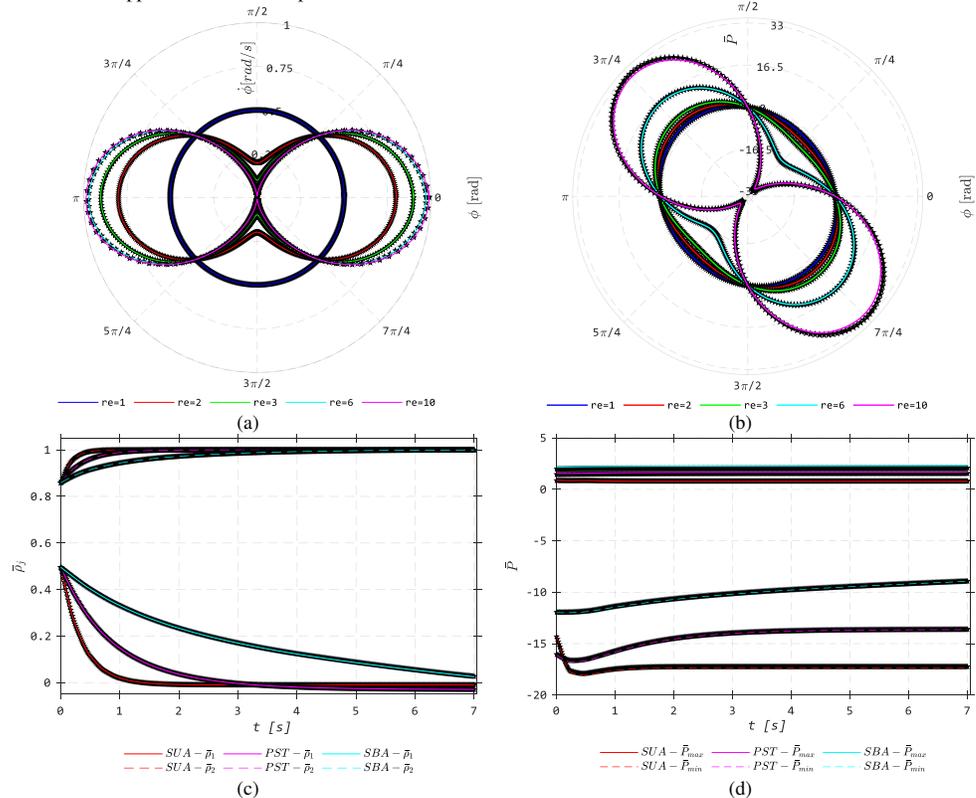

Figure 5[a]: FEA (colored lines) and Jeffery's results (black markers) of the evolution of the particle's (a) angular velocity, & (b) tip pressure, in simple shear flow for particle tumbling in the shear plane with different aspect ratios $1 \leq r_e \leq 10$; (c) orientation components, and (d) minimum (dashed) and maximum (continuous) surface pressure for particle with initial orientation, $\phi^0 = \pi/3$, $\theta^0 = 11\pi/24$, $\psi^0 = 0$ suspended in different combined flow types - SUA (red), PST (pink) and SBA (cyan) with $\dot{\gamma}/\dot{\varepsilon} = 1$.

## Results and Discussion

The Results and Discussion section is divided into two sub-sections. The first sub-section presents particles behavior (orientation dynamics and surface pressure distribution) in a Newtonian fluid, considering the various homogenous flows described above and the effect of geometric aspect ratio and particles initial orientation on the particles motion and evolution of the surface pressure. The subsequent sub-section presents in detail the effect of shear-thinning power-law fluid rheology on the particles behavior in the various combined homogenous flows and for different shear-to-extension rate ratio ($\dot{\gamma}/\dot{\varepsilon} = 1$ and 10). The section also presents the results of sensitivity studies on the influence of the ellipsoidal aspect ratio and initial particle orientation on the particles behavior in non-Newtonian simple shear flow.

---

[a] Results of the 3$^{rd}$ component of the particle's orientation vector (i.e. $\bar{\rho}_3$) is implicit given the normalization condition $\bar{\rho}_i \bar{\rho}_i = 1$.

**Particle Behavior in Newtonian Homogenous Flows**

For the investigation of the behaviour of single rigid spheroidal particle suspended in Newtonian homogenous flows, Jeffery's equations are sufficient and computationally more efficient than our numerical solutions. The basic homogenous flows discussed in the methodology section above that consider various combinations of stretching and shearing rate are expected in polymer composite melt flow applications such as material extrusion/deposition additive manufacturing (see e.g., Awenlimobor et al.,[63]). In all Newtonian flow analyses considered here, we employ an aspect ratio of $r_e = 6$, a viscosity of $\mu_1 = 1 \, Pa \cdot s$ and a shear rate of $\dot{\gamma} = 1 \, s^{-1}$ where applicable. The particle is initially oriented in the $\underline{X}_2$-direction (i.e. $\phi^0 = 0, \theta^0 = -\pi/2, \psi^0 = 0$) and rotates in the $\underline{X}_2 - \underline{X}_3$ shear plane.

Figure 6 shows the calculated particle in-plane angular velocity ($\dot{\phi}$) and particle tip pressure ($\bar{P}$) in the various homogenous flows for two cases of shear-to-extension rate ratio ($\dot{\gamma}/\dot{\varepsilon}$) where applicable. Here we use the overbar to indicate a dimensionless pressure as in eqns. 75 and 76. In the planar extensional flows (i.e. UA, BA, & TA flows), we observe an absence of particle motion, however, the particle begins to rotate with the introduction of a non-zero shear velocity gradient component (cf. Figure 6a). In the extension-shear SUA flow (i.e., $\dot{\gamma}/\dot{\varepsilon} = 1$), the particle is initially accelerated by the combined action of the inward flow in the $\underline{X}_2$-direction and the shear flow in the $\underline{X}_2-\underline{X}_3$ plane. The particle eventually stalls at $\phi_s = 1.58$ rad as it aligns with the $\underline{X}_3$-direction due to the applied stretching and relatively low shear rate. In the PST flow case, there is no flow in the $\underline{X}_2$-direction that influences the initial particle motion, however the inflow in the $\underline{X}_1$-direction keeps the particle motion in the $\underline{X}_2 - \underline{X}_3$ shear plane. Like the SUA flow case, the applied stretching and relatively high extensional dominance causes the particle to stall at $\phi_s = 1.60$ rad as it turns to align in the $\underline{X}_3$-direction. The SUA and PST mixed mode flow types are asymmetric in the $\underline{X}_2 - \underline{X}_3$ plane. In the SBA flow regime, the inward flow in the $\underline{X}_1$-direction prevents out-of-plane motion of the particle, and there is no provision for preferential orientation in the $\underline{X}_2 - \underline{X}_3$ plane due to uniform stretching in the $\underline{X}_2 - \underline{X}_3$-shear plane. As a result, the particle tumbles continuously. The STA and SS flow types are essentially similar in terms of their influence on the particle's behavior. The only difference observed between these flow types is in the calculated particle tip pressure. At the onset of particle motion at $\phi^0 = 0$ the net pressure at the particle tip is zero ($\bar{P} = 0$) for cases with no net flow in the $\underline{X}_2$-direction. However, the particle tip has a net positive pressure ($\bar{P} = +31.4$) for the UA/SUA flows due to the inflow in the $\underline{X}_2$-direction, and the outflow in the $\underline{X}_2$-direction creates a net negative pressure on the particles tip ($\bar{P} = -31.4$) for the BA/SBA cases. As the shear flow induces particle rotation, the tip pressure drops gradually until it reaches a minimum, at which point the particles orientation coincides with a principal flow direction (cf. Figure 6b).

In the event where the particle does not stall, the pressure on the particle tip fluctuates between its minimum and maximum limits at locations where its orientations coincides with the principal flow directions. For the axisymmetric flows, the particle tip pressure extremes occur at $\phi = \pm \pi/4$, while for the SUA asymmetric flow (i.e., $\dot{\gamma}/\dot{\varepsilon} = 1$), this occurs at $\phi = +1.41$ rad. Alternatively, for the PST asymmetric flow, the pressure extreme occurs at $\phi = +1.18$ rad. Cessation of the particles motion under the combined SUA and PST flow conditions is lifted once the conditions of eqn. 33 are violated, i.e. when $\dot{\gamma}/\dot{\varepsilon} \geq 3\kappa/\sqrt{1-\kappa^2}$ for the SUA flow condition and $\dot{\gamma}/\dot{\varepsilon} \geq \kappa/\sqrt{1-\kappa^2}$ for the PST flow conditions. In the current study where we assumed $\kappa = .9459$, the particle does not stall when $\dot{\gamma}/\dot{\varepsilon} \geq 8.75$ for SUA flow condition and when $\dot{\gamma}/\dot{\varepsilon} \geq 2.92$ for the PST flow condition. With increased shear strain rate (i.e., for $\dot{\gamma}/\dot{\varepsilon} = 10$), the particle rotates periodically for all combined flow conditions (cf. Figure 6c). Since $\dot{\varepsilon}_2 = \dot{\varepsilon}_3 = \dot{\varepsilon}$, for the axisymmetric combined flow cases, the particle does not stall regardless of the magnitude of $\dot{\gamma}/\dot{\varepsilon}$. One exception is seen for ellipsoidal particles with small but finite thickness such as in the case of a thin rod when $\kappa \to 1$ or in the case of a circular disc when $\kappa \to 0$, both of which are degenerate cases as described by Jeffery[15]. As the shear rate increases, the asymmetric flows becomes more symmetrical and the particle's surface pressure magnitudes are increased (cf. Figure 6d). Additionally, increased shear rate also moves the orientation where tip pressure extremes occurs (i.e. at the point where its coincides with the principal flow directions). For example, in the SUA flow case, the orientation where pressure extremes occurs are at $\phi = -0.640, +0.931$ rad while the same occurs at $\phi = -0.736, +0.835$ rad for the PST flow case.

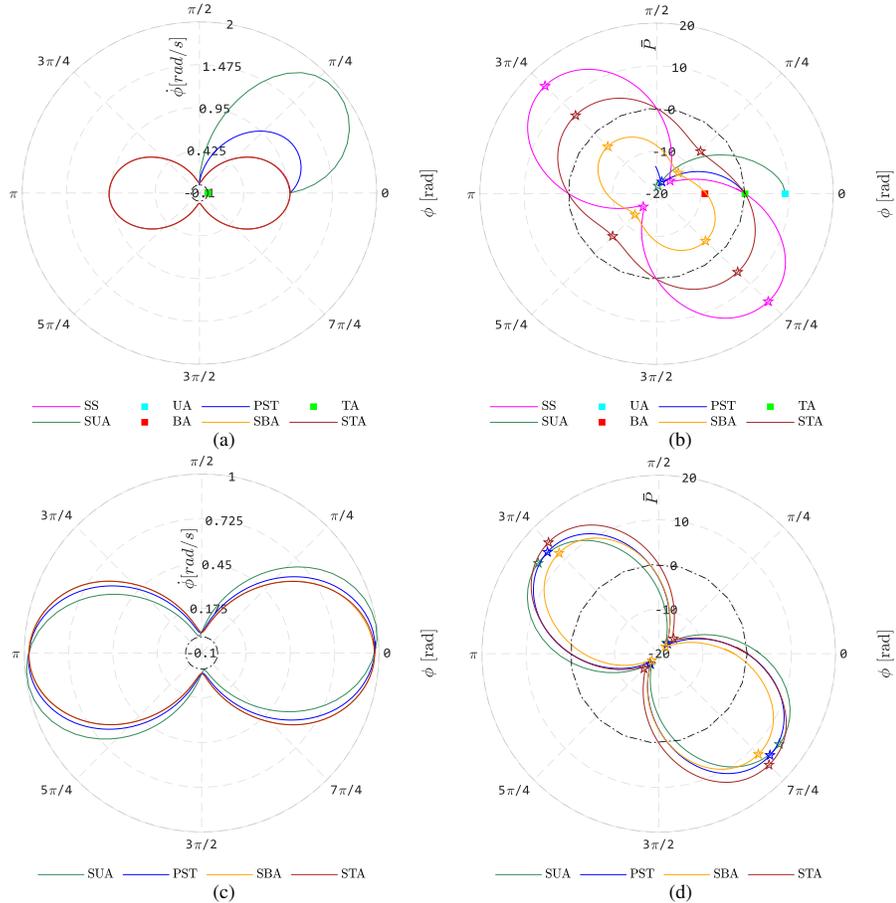

Figure 6: Polar plot of the evolution of the particle's (a) precession $\dot{\gamma}/\dot{\varepsilon} = 1$ (where applicable) (b) tip pressure $\dot{\gamma}/\dot{\varepsilon} = 1$ (c) precession $\dot{\gamma}/\dot{\varepsilon} = 10$ (d) tip pressure $\dot{\gamma}/\dot{\varepsilon} = 10$ for particle in the various homogenous flow types. In all cases, $\dot{\gamma} = 1\ s^{-1}, \mu_1 = 1\ Pa \cdot s$.

Particle motion analyses show that cessation of the rotation depends on the value of $\dot{\gamma}/\dot{\varepsilon}$, i.e. for the SUA and PST flows as shown in Figure 7. The tumbling period is seen to asymptote from either direction to the orientation where conditions for the onset of particle stall is satisfied which is seen to occur at a limit stall angle of approximately $\phi_p = 1.72$ rad. To the left of the red-dashed vertical limit lines in Figure 7a, or beneath the red-dashed horizontal line in Figure 7b, defining the asymptote events, the particle would stall, however the reverse situation is expected beyond these limits.

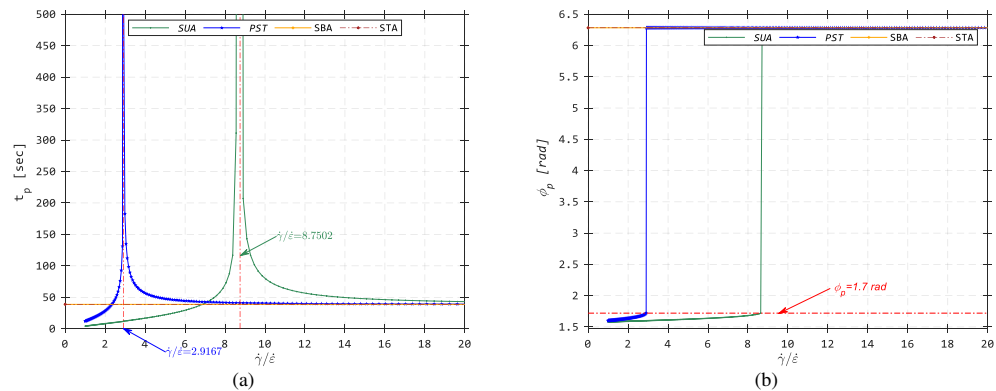

Figure 7: Particle motion analysis (a) particle tumbling period (or stall time where applicable) and (b) corresponding particle rotation angle, for different shear dominance factor $\dot{\gamma}/\dot{\varepsilon}$ and for the for the combined homogenous flow conditions.

**Effect of Geometric Aspect Ratio and Shear Rate**

In a 2D Newtonian study by Awenlimobor et. al.[63], various factors were shown to influence the peak pressure extreme on the surface of a particle suspended in Newtonian purely viscous simple shear flow including the fluid viscosity $\mu_1$, the magnitude of the shear rate $\dot{\gamma}$, and the particle aspect ratio $r_e$. The current investigation explores the 3D particle behavior in Newtonian purely viscous flow using Jeffery's equations. For a given aspect ratio, the net pressure $p - p_0$, computed from eqn. 1 is seen to have a linear dependence on the Newtonian viscosity $\mu_1$ and shear rate $\dot{\gamma}$, i.e. $(p - p_0)/\mu_1\dot{\gamma}$ is constant. However, as $r_e$ increases, so does the extreme tip pressure. Figure 5b shows that the particle's tip pressure magnitude is proportional to the $r_e$ of the ellipsoidal particle, which is likely due to the increased particle length, the reduced particle tip curvature which occurs as $r_e$ is increased, or both. From eqns. 18 and 19, it can be shown that the particle's tip pressure extremes occur at an orientation angle of $\phi = \pm \pi/4$ when the angular velocity $\dot{\phi} = \dot{\gamma}/2$ which also corresponds to the principal flow directions for simple shear flow. Further, at the position where the particle's precession approaches extremum at $\phi = n\pi/2, |n| \geq 0$, the particles tip pressure goes to zero irrespective of the geometric aspect ratio. Figure 8 shows the pressure distribution on the surface of rigids spheroidal particles at the location of orbital minimum surface pressure extreme for different aspect ratios and for particle motion in the plane of shear flow. It is evident that the minimum pressure on the particles surface occurs at the particle tips and the pressure peak magnitudes increases with the geometric aspect ratio.

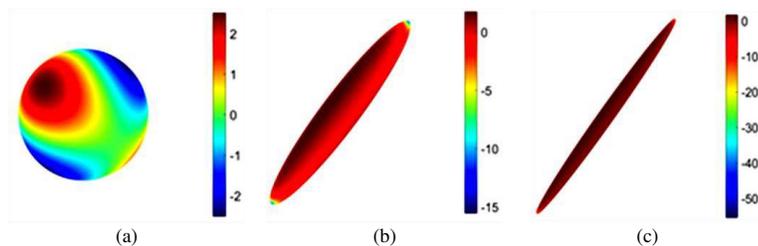

Figure 8: Pressure Distribution around the particle surface at the point of minimum peak pressure occurrence ($\phi = \pi/4$) for different aspect ratio (a) $r_e = 1$ (b) $r_e = 6$ (c) $r_e = 15$

With increased ellipsoidal aspect ratio, the curvature radius at the particle's tip reduces. It is important to understand the relation of the tip pressure magnitude with the tip geometry (i.e. the curvature radius, $r_\kappa = 1/r_e$) and with the relative positioning of the tip in the constant velocity gradient flow-field (defined by the particles geometric parameter,

19/40

$\kappa$). Figure 9a shows the relationship between the spheroidal orbital minimum tip pressure, $\bar{P}_{min,\kappa}$ normalized with respect to the spherical reference values, $\bar{P}_{min,0}$, (i.e. $\kappa = 0$) and the curvature radius for a prolate spheroid with unity minor axis length. This relationship obtained through a typical curve fitting procedure can be represented by eqn. 76. The Newtonian orbital minimum tip pressure ratio is seen to decrease exponentially with increasing tip curvature radius as

$$\bar{P}_{min,\kappa}/\bar{P}_{min,0} = 0.63 + 0.39 r_\kappa^{-1.53} - 4.81 \exp(14.47 r_\kappa) \qquad 76$$

Alternatively, the Newtonian orbital minimum tip pressure ratio can be represented in terms of the geometric parameter $\kappa$ as shown in Figure 9b and can be written as

$$\bar{P}_{min,\kappa}/\bar{P}_{min,0} = 1.87\kappa + 10.74\kappa^{19.56} + 0.82 \exp(4.54\kappa^{56.62}) \qquad 77$$

Figure 9b shows that as $\kappa$ tends to unity approaching a slender rod, the particle tip orbital minimum pressure goes to infinity. Note that the mean aspect ratio of short fiber fillers experimentally measured in 13% CF/ABS large scale EDAM printed bead were found to be about $r_e = 45$, $\kappa = 0.999$[78,79], that would theoretically yield high pressure spikes at the particle tips in the polymer suspension during polymer composite processing based on Jeffery's model assumption, which have been suggested by Awenlimobor et al.[63] to be potentially responsible for micro-void nucleation at the fiber tips.

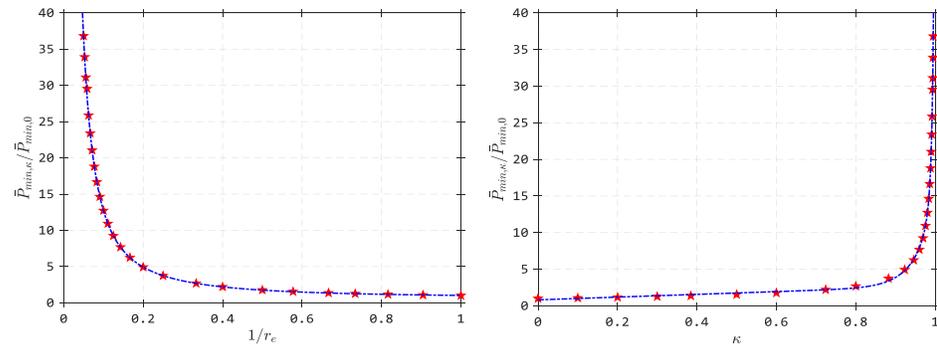

Figure 9: Relationship between the particle's orbital minimum pressure normalized with respect to the minimum surface pressure on a sphere in Newtonian fluid flow as a function of (a) radius of curvature ($r_\kappa$), and (b) geometric parameter $\kappa$. Results are shown for particle tumbling in simple shear flow with $\mu_1 = 1\ Pa \cdot s$ and $\dot{\gamma} = 1 s^{-1}$.

**Effect of Initial Particle Orientation**

In Figure 10a, we present the particle's motion in simple shear flow for various initial particle azimuth angle $\theta_0 = 2\pi/24 \leq \theta \leq 11\pi/24$ ($\phi_0 = 0$) based on Jeffery's solution given above. As expected, the particle's motion is periodic, and the period is same for all orbits. The orbit becomes narrower as we increase the initial out-of-plane orientation angle which reduces the effective aspect ratio (seen as that projected to the shear plane), resulting in lower peak pressure extremes. Figure 10b shows that the angle at which the particle pressure extreme occurs shifts as the particle is oriented further out of the shear plane. Eventually, setting the initial out of plane orientation to zero would lead to the particle spinning about its axis in a log-rolling position with near-zero surface pressure due to negligible disturbance velocity. The phase diagrams (cf. Figure 10c and d) reveals a symmetric behavior in particle dynamics. As the particle moves further out of plane (i.e. $\varsigma \to 0$), the location of the tip pressure extremes converges towards the location of minimum precession at $\phi = \pm \pi/2$, but as the particle moves towards the shear-plane, the pressure extreme locations coincide with the direction of the principal axis of the flow ($\phi = \pm \pi/4$). Figure 11 shows the particle's configuration at the location of minimum particle tip pressure along select Jeffery's orbits with various initial azimuth angle $\theta_0$. For the particle tumbling in the shear plane of the flow ($\theta_0 = -\pi/2$) we see that the particle's orientation coincides with the principal direction of the flow ($\phi = \pi/4$) but as it moves further out of plane, the peak pressure location moves closer towards the upper limit of azimuthal inclination for each orbit (i.e. $\phi \to \pi/2$).

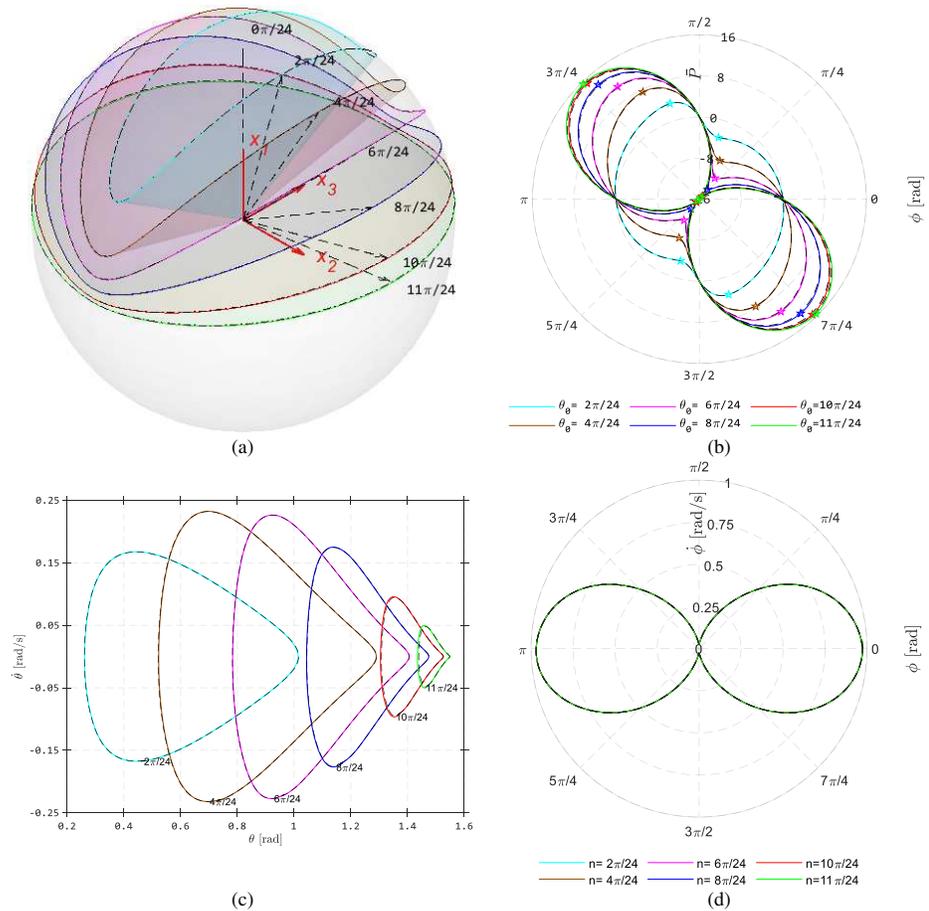

(a) (b) (c) (d)

Figure 10[b]: Results for different initial particle orientation showing (a) Jeffery's orbits (b) particle tip pressure evolution where the asterisk (*) indicates location of the tip pressure extreme (c) phase diagram of azimuth angle $\theta$ vs nutation $\dot{\theta}$ (d) polar plot of the precession $\dot{\phi}$ vs polar angle $\phi$. Results are shown for $-2\pi/24 \leq \theta_0 \leq -12\pi/24$ and for simple shear flow with $\mu_1 = 1\,Pa \cdot s$ and $\dot{\gamma} = 1s^{-1}, r_e = 6$.

---

[b] The results presented in Figure 14(a)-(d) are also validated with both FEA and Jeffery's analytical calculation. The black dashed lines are results obtained from Jeffery's equation and the continuous colored lines are results from FEA computations.

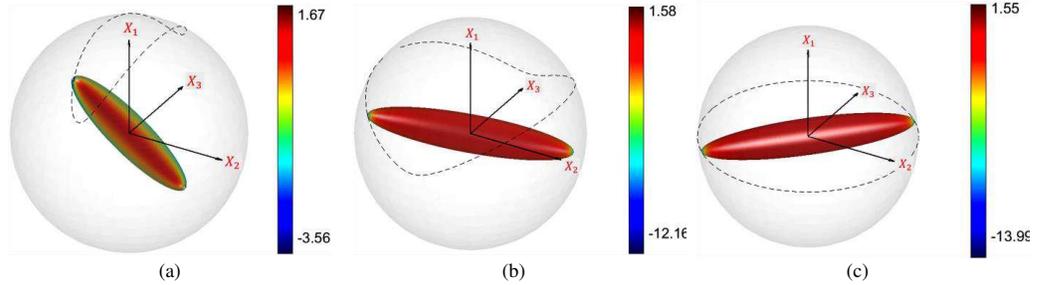

Figure 11: Spatial configuration of the particle at the point of minimum pressure occurrence and for various initial azimuthal angle $\theta_0$ of (a) $\theta_0 = 2\pi/24$, (a) $\theta_0 = 8\pi/24$, (a) $\theta_0 = 12\pi/24$. Results are shown for $\mu_1 = 1\ Pa \cdot s$ and $\dot{\gamma} = 1s^{-1}$, $r_e = 6$.

Figure 12a shows a nearly linear relationship between the particle's orbital minimum tip pressure and the polar angle location along the corresponding Jeffery's orbits. As noted above, when the particle is tumbling in shear plane (i.e., $\varsigma = +\infty$), the location of the particle's surface extreme pressure coincides with the ellipsoidal tip location. However, as the particle becomes oriented more out-of-plane (i.e. $\varsigma \to 0$), the location of minimum pressure on the particle surface at the orientation of peak pressure occurrence is slightly shifted away from the tip down the leeward side trailing the flow. Figure 12b shows the difference between the minimum pressure on the fibers surface and tip pressure ($\delta \bar{P}$) at the instant when the peak occurs along Jeffery's orbit. The result shows that a higher initial out of plane orientation leads to greater deviation of the fiber tip pressure from its surface pressure extreme magnitude.

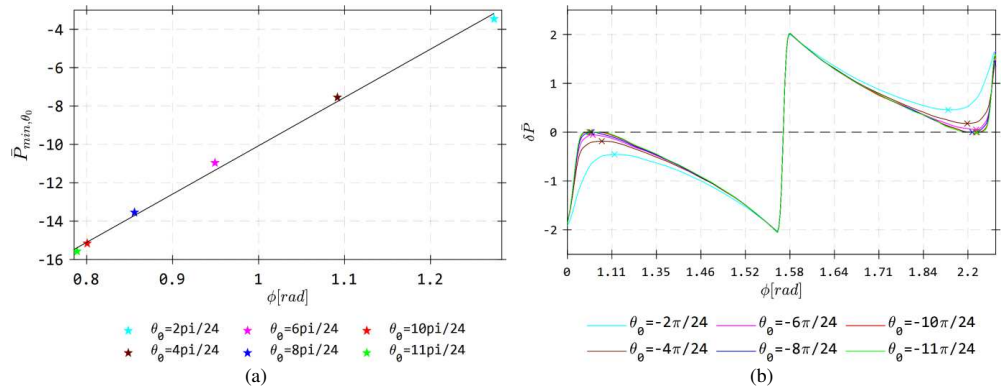

Figure 12: Tip pressure results (a) Orbital minimum particle tip pressure versus polar angle, and (b) difference in the instantaneous particle tip pressure and actual surface pressure extremum, for different Jeffery's orbit and for $\mu_1 = 1\ Pa \cdot s$ and $\dot{\gamma} = 1s^{-1}$, $r_e = 6$.

The particle orbital maximum nutation $\dot{\theta}$ itself peaks at a Jeffery's orbit that passes through $(\phi, \theta = \pm \pi/4)$ irrespective of the aspect ratio. In Figure 13a, the continuous lines trace the paths of orbital maximum nutation across the degenerate spectrum of Jeffery's orbit for different aspect ratios, and the dashes lines are the Jeffery's orbit that cuts across the location of peak nutation for different ellipsoidal aspect ratios. From Figure 13b, the peak nutation across the spectrum of Jeffery's orbit is observed to increase with the aspect ratio and approaches the critical value at $\dot{\theta} = \dot{\gamma}/4$.

22/40

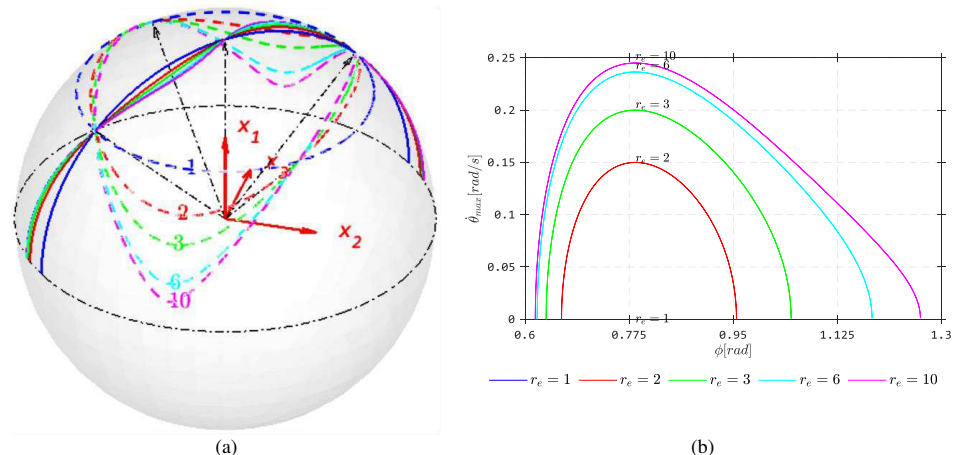

(a)  (b)

Figure 13: Out-of-plane Jeffery orbits (a) the path of orbital maximum nutation across degenerate spectrum of Jeffery's orbit for aspect ratios of 1, 2, 3, 6, and 10 (continuous lines) and critical Jeffery's orbit at which the orbital maximum nutation attains peak magnitude for the same aspect ratios (dashed lines). (b) phase plot of the orbital maximum nutation across degenerate spectrum of Jeffery's orbit for different aspect ratios.

**Particle Motion in Non-Newtonian Homogenous Flows**

The results presented above focused on a single rigid ellipsoidal particle in various combined extensional and shear Newtonian homogenous flows that are considered typical of those in an EDAM nozzle during polymer composite processing. It is well understood, however, that thermo-plastic polymer materials are inherently non-Newtonian. Moreover, the addition of filler reinforcements to polymers are known to increase the melt viscosity and the shear-thinning fluid behavior in the nozzle. Additionally, high shear regions of complex flows such as the lubrication zone near the screw edge or regions of flow acceleration near the nozzle are known to result in flow segregation of highly shear-thinning polymer melt suspension into resin lean highly viscous domains and resin rich low-viscosity domains. As such understanding the particle behavior in shear-thinning fluid within various flow regimes is important in understanding microstructural development within polymer composite beads. The sections to follow present results obtained with the nonlinear FEA modeling approach presented above which considers a non-Newtonian shear-thinning power-law fluid rheology.

Prior research on the response of a single particle suspended in a 2D viscous flow performed by Awenlimobor et al[64] showed that although the shear-thinning rheology has no impact on the particles motion, the particle surface pressure extremes are reduced with decreasing power-law index. Here we consider the response of a single 3D ellipsoidal particle in simple homogeneous power-law fluid flows computed using the FEA method described above. The results presented in Figure 14 are for an ellipsoid with geometric ratio $r_e = 6$ rotating in a power-law fluid with a flow shear rate of $\dot{\gamma} = 1\ s^{-1}$ and power-law indices ranging from 0.2 to 1.0.

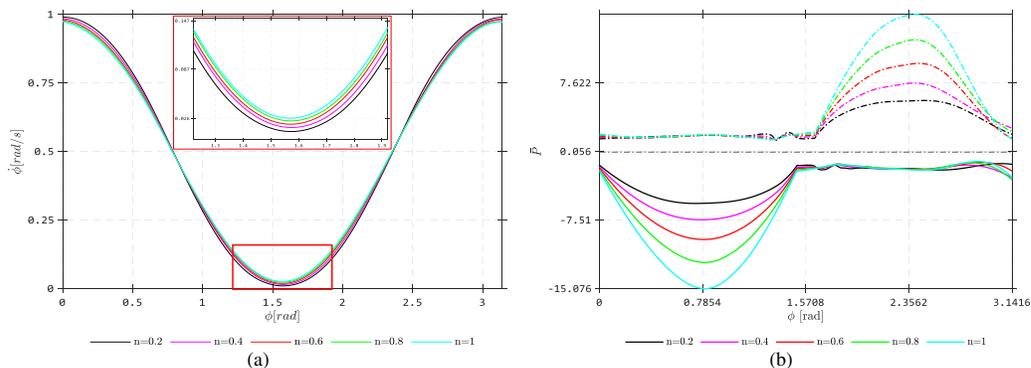

Figure 14: FEA computed shear-thinning response of (a) particle polar angle $\phi$ vs precession $\dot{\phi}$ and (b) surface pressure extremes for particle motion in simple shear flow. Results are shown, for $r_e = 6$, $0.2 \leq n \leq 0.8$, $m = 1 \, Pa \cdot s^n$, $\dot{\gamma} = 1s^{-1}$ and $\phi^0 = 0, \theta^0 = -\pi/2, \psi^0 = 0$.

Figure 14a shows that the shear-thinning behavior has a slight influence on the particle's dynamic motion as reduction in the power-law index slows down the particle. The limits of the particle's in-plane angular velocity are observed to increase with increasing power-law index. Further, Figure 14b shows that the particle surface pressure extremes increase with decreased shear-thinning. Additionally, it is interesting to note that even though the orbit formed from particle tumbling in the shear-plane appears to exhibit little noticeable difference due to shear-thinning, Figure 15a shows that the tumbling period significantly increases with increasing shear-thinning.

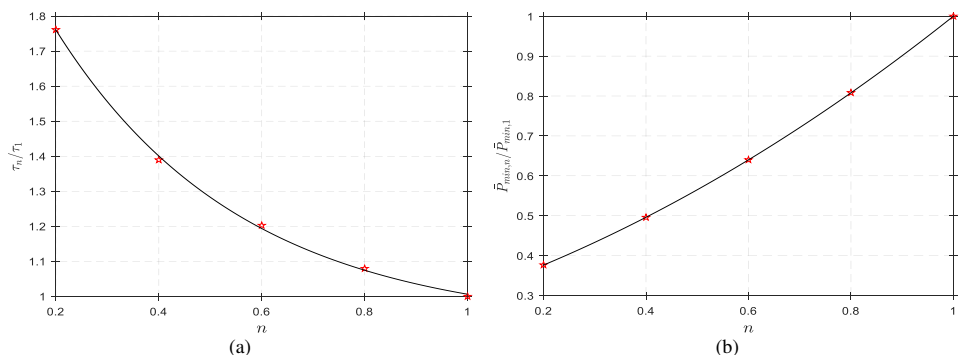

Figure 15: Non-Newtonian to Newtonian ratio of the (a) particle's in-plane tumbling period (b) Orbital minimum particle tip pressure. Results are shown for $r_e = 6$, $0.2 \leq n \leq 0.8$, $m = 1 \, Pa \cdot s^n$, $\dot{\gamma} = 1s^{-1}$ and $\phi^0 = 0, \theta^0 = -\pi/2, \psi^0 = 0$.

The relationship between the particle tumbling period $\tau_n$ and the power-law index $n$ under simple shear flow conditions was determined through a typical curve fitting procedure to follow

$$\tau_n = \tau_1(0.9135 + 1.4724 e^{-2.7645n}) \tag{78}$$

where $\tau_n$ is the tumbling period in a shear-thinning fluid with power-law index $n$ and $\tau_1$ is the particle tumbling period for the Newtonian case, i.e. when $n = 1$. Figure 15b shows that the orbital minimum particle tip pressure has a quadratic variation with the flow behavior index as described as

$$\bar{P}_{min,n} = \bar{P}_{min,1}(0.28 + 0.42n + 0.30n^2) \tag{79}$$

which implies that the shear-thinning effect on particle pressure distribution can be interpreted as having the same effect as would a modification of the Newtonian viscosity, agreeing with the findings of Ji et al[53] and Awenlimobor et al[64].

Figure 16 shows the pressure field around the ellipsoidal particle at various instants during the particle tumbling motion in the plane of the shear flow. The contours show an intensification of the pressure on the particle surface as the power-law index increases from $n = 0.2$ to $n = 1.0$. The pressure intensification is observed to be higher at orientations of peak orbital pressure extreme magnitudes (i.e. at $\phi = \pm \pi/4$). These observations can be explained from the plot of the disturbance in the velocity $\dot{X}_i^{d\,34}$ around the surface of the particle due to the particles motion defined as the difference between the flow-field velocity and free stream velocity, i.e. $\dot{X}_i^d = \dot{X}_i - \dot{X}_i^\infty$ (cf. Figure 17). We observe a higher magnitude of the velocity disturbance around same location on the particles surface where pressure extremes are observed to occur (i.e. at the particle tips). Likewise, the intensity of the disturbance is seen to increase with increasing power-law index and the magnification is higher at critical orientation angles where the orbital peak pressure extremes occur during alignment with the principal flow directions (i.e. at $\phi = \pm \pi/4$). The lower pressure intensities are thus a result of lower disturbance in the velocity field around the particle caused by the deceleration of the particles motion in the shear-thinning fluid.

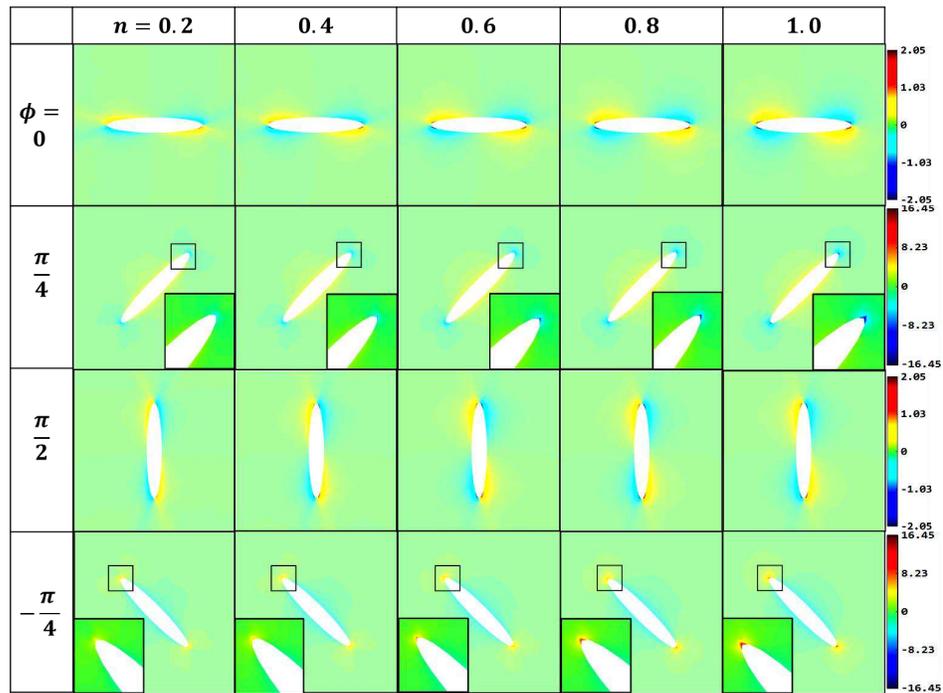

Figure 16: Mid - sectional plot of the pressure distribution around the ellipsoidal particle for at different instants during the particle's in-plane tumbling motion ($\phi = 0, \pi/4, \pi/4, \pi/4$) and for different power-law indices ($0.2 \leq n \leq 0.8$). Results are shown for $r_e = 6, m = 1\,Pa \cdot s^n, \dot{\gamma} = 1s^{-1}$ and $\phi^0 = 0, \theta^0 = -\pi/2, \psi^0 = 0$.

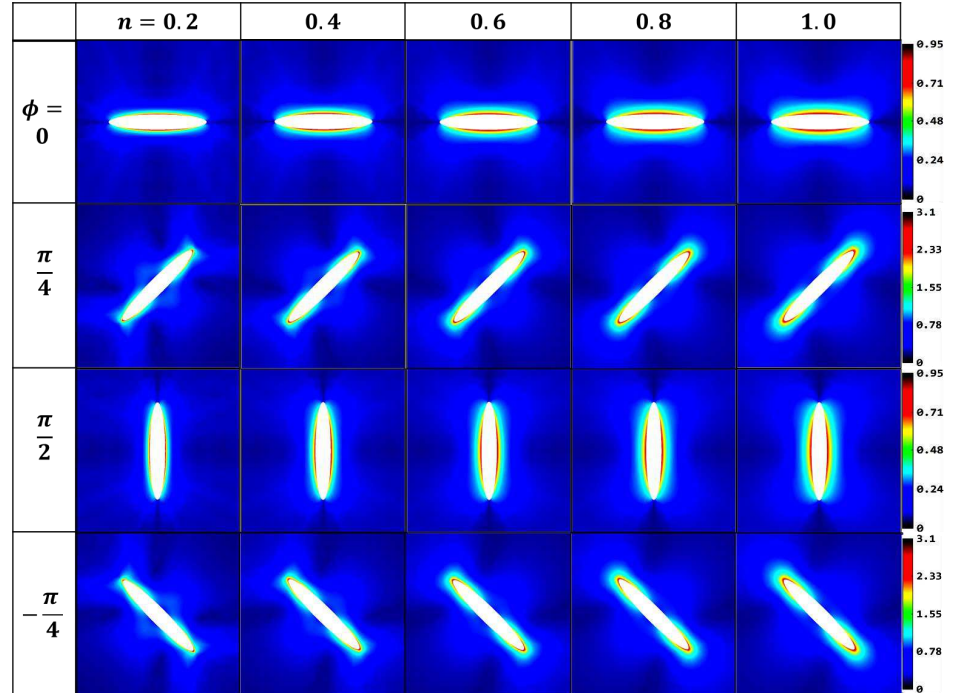

Figure 17: : Mid - sectional plot of the disturbance velocity around the ellipsoidal particle for at different instants during the particle's in-plane tumbling motion ($\phi = 0, \pi/4, \pi/4, \pi/4$) and for different power-law indices ($0.2 \leq n \leq 0.8$). Results are shown for $r_e = 6$, $m = 1\, Pa \cdot s^n$, $\dot{\gamma} = 1 s^{-1}$ and $\phi^0 = 0, \theta^0 = -\pi/2, \psi^0 = 0$.

Figure 18a-d shows the computed results of the single rigid ellipsoidal particle in combined shear and uniaxial extension (SUA) flow type with a power-law index $n$ ranging from 0.2 to 1 while considering two shear-extension rate ratios (i.e., $\dot{\gamma}/\dot{\varepsilon} = 1$ and 10). Figure 18a and Table II shows that the particle stalls in the SUA flow with $\dot{\gamma}/\dot{\varepsilon} = 1$) and the shear-thinning fluid behavior slightly increases particle rotation speed and shortens the trajectory which is evident from the slight reduction in the time to particle stall and the stall angle with decreasing power-law index. Figure 18b shows that the shear-thinning fluid reduces the magnitude of the particle surface pressure extremes in the SUA flow, however, the shear-thinning rheology does not affect the orbital angle location where the minimum peak magnitude pressure occurs (i.e. at $\phi = +1.41$ rad).

| $n$ | 0.2 | 0.4 | 0.6 | 0.8 | 1.0 |
|---|---|---|---|---|---|
| $\tau_s$ | 3.922 | 3.982 | 4.012 | 4.032 | 4.032 |
| $\phi_s$ | 1.574 | 1.577 | 1.579 | 1.580 | 1.580 |

Table II: Particle stall time $\tau_s$ and particle stall angle $\phi_s$ for single ellipsoidal particle motion in SUA shear-thinning flow for different flow behavior index $0.2 \leq n \leq 1.0$ with $m = 1\, Pa \cdot s^n$, $\dot{\gamma} = 1 s^{-1}$ and $\dot{\gamma}/\dot{\varepsilon} = 1$.

In the shear dominant flow condition when $\dot{\gamma}/\dot{\varepsilon} = 10$, the particle tumbles periodically under slightly non-Newtonian rheological fluid behavior ($n \geq 0.8$), however further reduction in the power-law index ($n < 0.8$) causes the particle to eventually stall in a preferred orientation along the direction of stretching (cf. Figure 18c). This implies that the conditions for particle stall in a shear-thinning fluid is dependent on the competing influence of the shear-extensional rate factor and the intensity of the shear-thinning fluid behavior. Table III shows that the particle stall time ($\tau_s$) and stall angle ($\phi_s$) when $n < 0.8$, and half period ($\tau_n^{0.5}$) for the cases where the particle tumbles periodically (i.e. when $n \geq 0.8$). As expected, at the location of the orbital extreme pressure magnitude where the particle orientation coincides with the principal flow direction (at $\phi = +0.931, +2.502$ rad), the surface extreme pressure magnitudes are observed to decrease with the intensity of the shear-thinning fluid rheology (cf. Figure 18&d). The pressure fluctuations on the particle's tip as it tumbles continuously in the shear dominant flow or the local pressure that subsist at particle's tip as it stalls in the extension dominant flow condition are important in understanding the final microstructural formations within printed polymer composite beads[63].

| $n$ | 0.2 | 0.4 | 0.6 | 0.8 | 1.0 |
|---|---|---|---|---|---|
| $\tau_n^{0.5}$ (or $\tau_s$) | 31.930 | 39.777 | 65.146 | 59.070 | 40.156 |
| $\phi_s$ | 1.607 | 1.644 | 1.689 | - | - |

Table III: Half-period/stall time (where applicable) $\tau_s$ and stall angle $\phi_s$ (where applicable) for single ellipsoidal particle motion in SUA shear-thinning flow for different flow behavior index $0.2 \leq n \leq 1.0$ with $m = 1\ Pa \cdot s^n$, $\dot{\gamma} = 1 s^{-1}$ and $\dot{\gamma}/\dot{\varepsilon} = 10$.

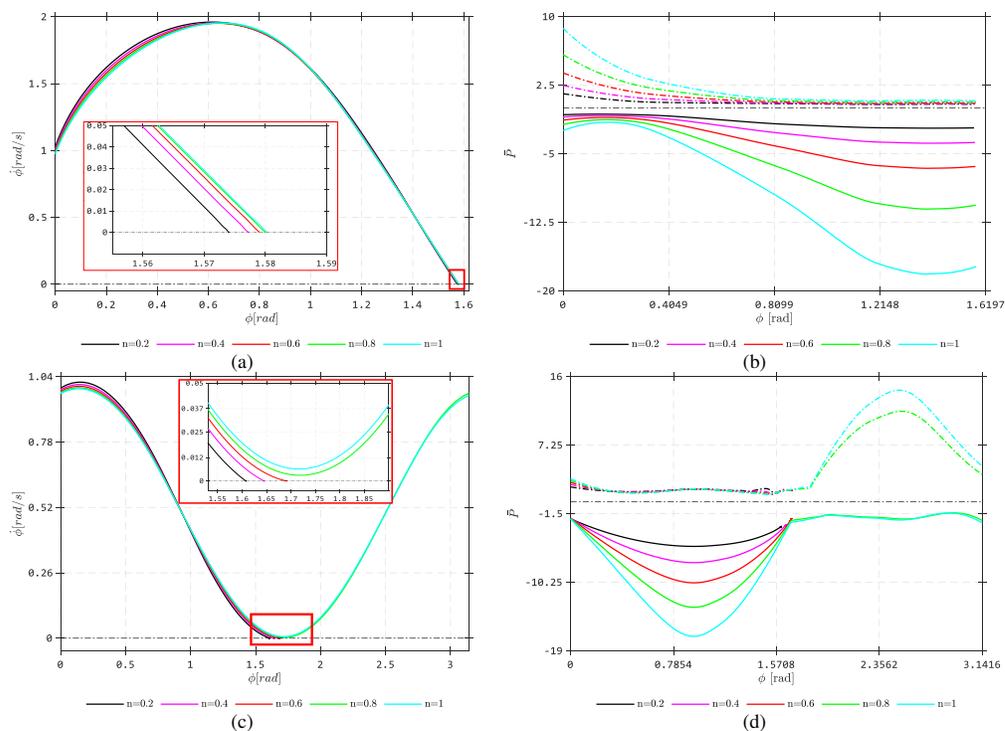

Figure 18: Phase diagram of the particles polar angle $\phi$ vs (a) precession $\dot{\phi}$ - $\dot{\gamma}/\dot{\varepsilon} = 1$ (b) surface pressure maximum (dashed) and minimum (continuous) - $\dot{\gamma}/\dot{\varepsilon} = 1$ and (c) precession $\dot{\phi}$ - $\dot{\gamma}/\dot{\varepsilon} = 10$, (d) surface pressure maximum (dashed) and minimum (continuous) - $\dot{\gamma}/\dot{\varepsilon} = 10$, for particle motion in combined shear and uniaxial extension (SUA) flow. Results are shown, for $0.2 \leq n \leq 0.8$, $m = 1\ Pa \cdot s^n$, $\dot{\gamma} = 1 s^{-1}$ and $\phi^0 = 0, \theta^0 = -\pi/2, \psi^0 = 0$.

In the combined shearing/planar stretching (PST) flow, the shear-thinning fluid rheology does not deter the particle's acquiescence into preferred orientation state under the extension-rate dominant flow condition (i.e. $\dot{\gamma}/\dot{\varepsilon} = 1$). However, the shear-thinning is observed to decelerate the particles motion, prolong the stall event and extend the particles trajectory to stall contrary to what was observed in the SUA flow. Figure 19a reveals a slight reduction in the peak in-plane angular velocity with decreasing power-law index and Table IV shows that the stall time and stall angle both of which increase with increased shear-thinning. The particle tip pressure magnitudes are nonetheless observed

27/40

to decrease with increased shear-thinning as expected (cf. Figure 19b). The particle in-plane orientation at the location of orbital minimum surface pressure (i.e. at $\phi = +1.18$) is unaltered by the shear-thinning effect.

| $n$ | 0.2 | 0.4 | 0.6 | 0.8 | 1.0 |
|---|---|---|---|---|---|
| $\tau_s$ | 13.337 | 11.796 | 11.026 | 10.515 | 10.135 |
| $\phi_s$ | 1.676 | 1.644 | 1.625 | 1.611 | 1.600 |

Table IV: Particle stall time $\tau_s$ and particle stall angle $\phi_s$ for single ellipsoidal particle motion in PST shear-thinning flow for different flow behavior index $0.2 \leq n \leq 1.0$ with $m = 1\ Pa \cdot s^n$, $\dot{\gamma} = 1s^{-1}$ and $\dot{\gamma}/\dot{\varepsilon} = 1$.

The shear-thinning effect does not stall the particle under the shear-rate dominant condition (i.e. when $\dot{\gamma}/\dot{\varepsilon} = 10$) in the PST flow contrary to what was observed in the SUA flow. However, at the local minima of the particle's angular velocity evolution curve when its deceleration approaches zero (cf. Figure 19c), the increased shear-thinning effect is observed to further decelerate particle motion and bring it closer to stall condition. Table V shows that the particles tumbling period increases with decreasing power-law index indicating the deceleration of the particle rotation with increased shear-thinning. The sustained particle motion allows for continuous fluctuations between particle surface pressure extremes at the particle tip. As would be expected, the pressure magnitudes are observed to decrease with increased shear-thinning (cf. Figure 19d). Further, the in-plane orientation at the orbital location of particle surface tip pressure extremum (i.e. at $\phi = +0.835, +2.406\text{rad}$) is unaltered by the shear-thinning effect. The particle's tumbling period is likewise observed to increase with decreasing power-law index due to increased particle deceleration induced by the shear-thinning fluid rheology (cf. Table IV).

| $n$ | 0.2 | 0.4 | 0.6 | 0.8 | 1.0 |
|---|---|---|---|---|---|
| $\tau_n^{0.5}$ | 36.181 | 28.045 | 24.169 | 21.839 | 20.280 |

Table V: Half-period for single ellipsoidal particle motion in PST shear-thinning flow for different flow behavior index $0.2 \leq n \leq 1.0$ with $m = 1\ Pa \cdot s^n$, $\dot{\gamma} = 1s^{-1}$ and $\dot{\gamma}/\dot{\varepsilon} = 10$.

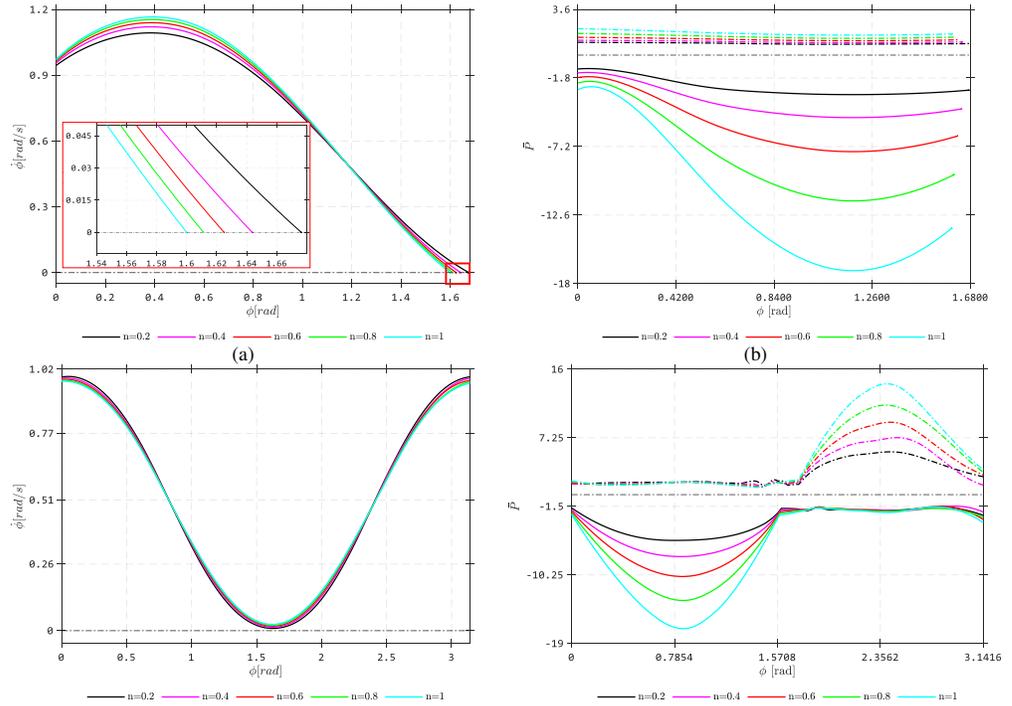

(c)                          (d)

Figure 19: Phase diagram of the particles polar angle $\phi$ vs (a) precession $\dot{\phi}$ - $\dot{\gamma}/\dot{\varepsilon} = 1$ (b) surface pressure maximum (dashed) and minimum (continuous) - $\dot{\gamma}/\dot{\varepsilon} = 1$ and (c) precession $\dot{\phi}$ - $\dot{\gamma}/\dot{\varepsilon} = 10$, (d) surface pressure maximum (dashed) and minimum (continuous) - $\dot{\gamma}/\dot{\varepsilon} = 10$, for particle motion in combined shear and planar stretching (PST) flow. Results are shown, for $0.2 \leq n \leq 0.8$, $m = 1\ Pa \cdot s^n$, $\dot{\gamma} = 1s^{-1}$ and $\phi^0 = 0, \theta^0 = -\pi/2, \psi^0 = 0$.

Under the balanced shear and bi-axial elongation (SBA) flow condition, inward flow normal to the shear plane coupled with uniform stretching along the shear plane promotes particle in-plane tumbling motion. Under this flow condition, the particle does not stall irrespective of the magnitude of the extension rate. However, while the increased shear-thinning is observed to accelerate the particles motion when $\dot{\gamma}/\dot{\varepsilon} = 1$, it is shown to slightly decelerate the particles motion under a higher shear rate i.e. $\dot{\gamma}/\dot{\varepsilon} = 10$ (cf. Figure 20a & c). When $\dot{\gamma}/\dot{\varepsilon} = 1$ the limits of particle in-plane angular velocity are observed to decrease with increased shear-thinning and vice versa when $\dot{\gamma}/\dot{\varepsilon} = 10$. The shear-thinning effect decreases the particle tumbling period when $\dot{\gamma}/\dot{\varepsilon} = 1$ and increases the period when $\dot{\gamma}/\dot{\varepsilon} = 10$ (cf. Table VI). Under a lower shear rate ($\dot{\gamma}/\dot{\varepsilon} = 1$), there are no noticeable peaks in the evolution of the particle maximum surface pressure, contrary to what is observed when $\dot{\gamma}/\dot{\varepsilon} = 10$. As would be expected, the particle surface pressure extremes are observed to decrease with increased shear-thinning and the location of orbital minimum surface pressure at $\phi = \pm \pi/4$ is unaffected by the shear-thinning rheology (cf. Figure 20b and d).

| $\tau_n^{0.5}$ | | $n$ | | | | |
|---|---|---|---|---|---|---|
| | | 0.2 | 0.4 | 0.6 | 0.8 | 1.0 |
| $\dot{\gamma}/\dot{\varepsilon}$ | 1 | 9.558 | 11.273 | 13.266 | 15.799 | 19.453 |
| | 10 | 25.265 | 22.650 | 21.155 | 20.138 | 19.423 |

Table VI: Half tumbling period $\tau_n^{0.5}$ for single ellipsoidal particle motion in SBA shear-thinning flow for different flow behavior index $0.2 \leq n \leq 1.0$ and different shear to extension rate ratio ($\dot{\gamma}/\dot{\varepsilon}$) with $m = 1\ Pa \cdot s^n$, $\dot{\gamma} = 1s^{-1}$.

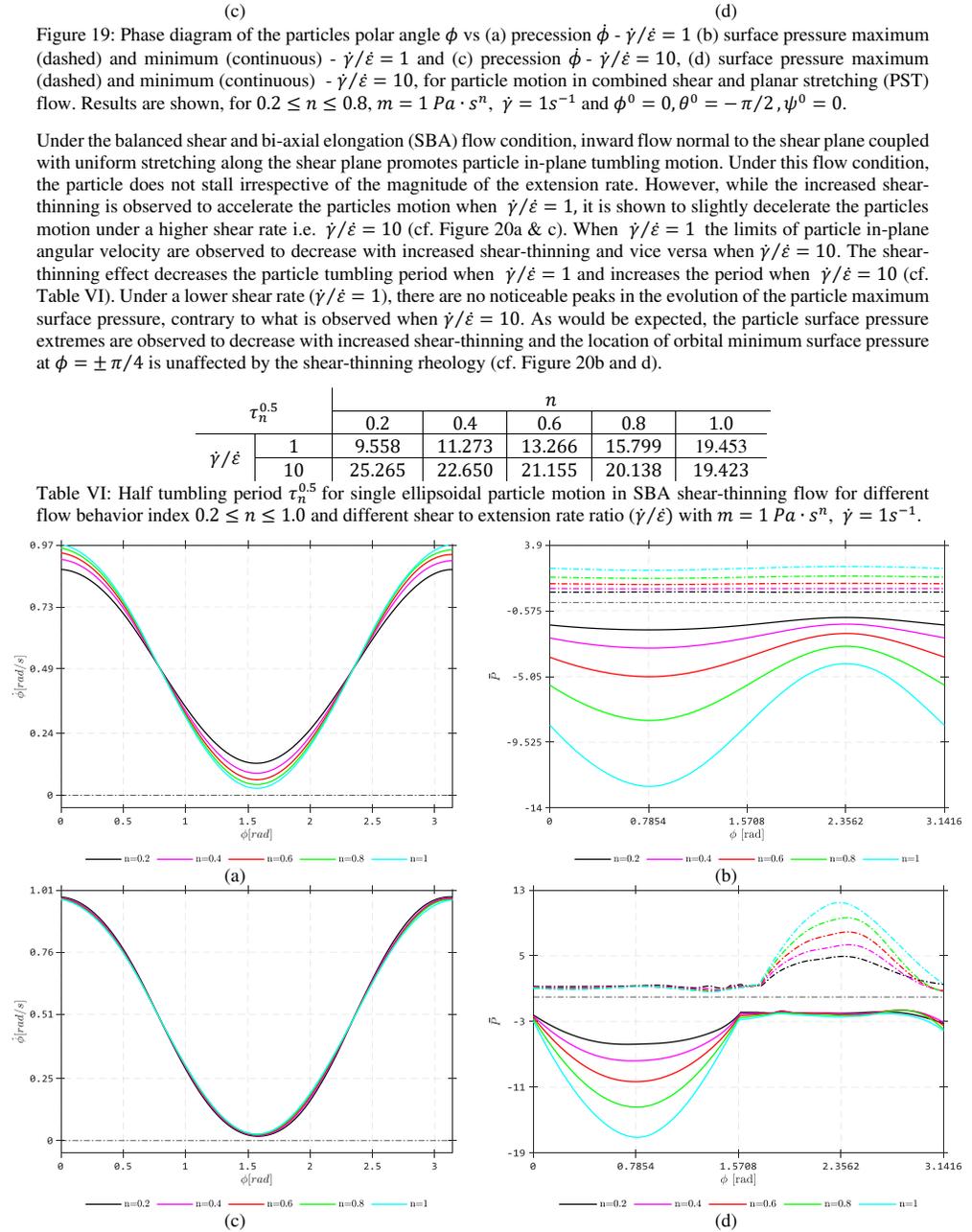

(a)                          (b)

(c)                          (d)

Figure 20: Phase plots of the particles polar angle $\phi$ vs (a) precession $\dot{\phi}$ - $\dot{\gamma}/\dot{\varepsilon} = 1$ (b) surface pressure maximum (dashed) and minimum (continuous) - $\dot{\gamma}/\dot{\varepsilon} = 1$ and (c) precession $\dot{\phi}$ - $\dot{\gamma}/\dot{\varepsilon} = 10$, (d) surface pressure maximum (dashed) and minimum (continuous) - $\dot{\gamma}/\dot{\varepsilon} = 10$, for particle motion in combined shear and biaxial extension (SBA) flow. Results are shown, for $0.2 \leq n \leq 0.8$, $m = 1\ Pa \cdot s^n$, $\dot{\gamma} = 1s^{-1}$ and $\phi^0 = 0, \theta^0 = -\pi/2, \psi^0 = 0$.

Observation of particle behavior in the flow types considered here as applied to polymer melt flow conditions during EDAM processing suggests that the shear-thinning effect increases the particle stall tendency closer to the EDAM nozzle center where a higher extension rate dominance is seen. Shear-thinning is seen here to have a similar effect as decreasing the shear-to-extension rate ($\dot{\gamma}/\dot{\varepsilon}$), thus shifting the boundaries of the extension dominant region outward (cf. Appendix B, Figure 24). Irrespective of the flow regime, the shear-thinning rheology reduces the pressure magnitude which has a similar effect to reducing the viscosity magnitude in a Newtonian fluid. Additionally, in high shear dominant flow regions of the EDAM nozzle, the shear-thinning effect is generally expected to slow down the particles motion, while close to the nozzle center, dominated by high extension-rate, the particle's stall event is expected to be promoted by shear-thinning effects.

**<u>Effect of Initial Particle Orientation</u>**

In earlier sections we showed that the pressure magnitudes on the surface of a particle suspended in a Newtonian simple shear flow reduces as the orbit constant $\varsigma$ (cf. eqn. 19,20) goes from $\varsigma = +\infty$ where the particle is tumbling in the shear plane to $\varsigma = 0$ where the particle is spinning about its axis perpendicular to the shear plane. It was also shown that the tumbling period was unaffected by Jeffery's orbit. The effect of shear-thinning rheology on the particle motion for various Jeffery orbits are presented in this section. We consider particle motion in simple shear flow with shear rate of $\dot{\gamma} = 1s^{-1}$ and for a GNF power-law fluid rheology with a power-law index of $n = 0.5$ and a consistency index of $m = 1\ Pa \cdot s^n$. The same geometric aspect ratio of $r_e = 6$ as was previously used is considered here.

Figure 21a shows that Jeffery's orbits are altered slightly by the shear-thinning fluid which occurs to a greater extent as the fiber is oriented further out of the shear plane (i.e., as $\varsigma = +\infty$ is moved to $\varsigma = 0$). The initial particle polar angle on a particular Newtonian Jeffery's orbit is observed to also modify the particle trajectory. Figure 21a and b also show that trajectory of the particle motion in an orbit with initial azimuth angle of $\theta^0 = 2\pi/24$ with two initial starting positions at the vertices of the Newtonian conical orbit. With an initial starting position at the vertex of the directrix of the Newtonian conical orbit on the major axis (at $\phi^0 = \pi/2$), the particle path is seen to dilate outwardly defined by the outer curve (dashed cyan line) from the Newtonian orbit (continuous black line). However, starting the particle from the co-vertex of the directrix of the Newtonian orbit on the minor axis (i.e. $\phi^0 = 0$), the orbit constricts inwardly defined by the inner curve (continuous cyan line). Both curves clearly illustrate the extent of deviation in the particle path from the Newtonian orbit and that for a given power-law index and set of flow parameters. The fluid shear-thinning is seen to influence the particles motion similar to elongating or shortening the particle, depending on the initial position on the orbit. This observed behavior is consistent with conclusions by Abtahi et al.[34].

The fluid shear-thinning is seen to have a more profound effect on the surface pressure of particles on Jeffery orbit closer to the shear plane ($\varsigma \to +\infty$) as compared to orbits farther out of plane (i.e. close to $\varsigma \to 0$). The net pressure drop ($\delta \bar{P}$) due to the shear-thinning effect is seen to be proportional to the magnitude of the particle surface pressure as shown in Figure 21c. Likewise, the net pressure drop of particle tip pressure is seen to depend on its initial starting position as is evident from the net pressure curves shown for each initial polar angle on the orbit farthest from the shear plane ($\theta^0 = 2\pi/24$), i.e. dashed cyan line for $\phi^0 = 0$ and continuous cyan line for $\phi^0 = \pi/2$.

As expected, the particle dynamics are also affected by the shear-thinning rheology. The envelope of the phase diagram of the particle's nutation (cf. Figure 21d) contract inwardly from the Newtonian envelope due to the shear-thinning effect irrespective of the initial position on the orbit. The shear-thinning rheology appears to have less effect on the particle's precession as the Jeffery's orbit is oriented further out of plane, i.e. when $\varsigma \to 0$ (cf. Figure 21e), however, this effect on the particle's nutation is more profound as $\varsigma \to 0$. Although, the particle's period of tumbling is independent on the Jeffery's orbit in Newtonian flow, the tumbling period is observed to be influenced by the Jeffery's orbit under shear-thinning flow conditions. Figure 22a shows the relationship between the tumbling period $\tau_{0.5}$ and the initial azimuth angle, $\theta_0$ for the particle motion in non-Newtonian power-law fluid, with flow behaviour index of $n = 0.5$. The relationship in Figure 22a can be described as

$$\tau_{0.5} = \tau_1 \left(1.2976 - 0.7358 e^{-3.8495 \theta_o}\right) \qquad 80$$

which has been obtained using a typical curve fitting procedure. Overall, the shear-thinning fluid rheology slows down a particle's motion which occurs to a greater degree as the tumbling orbit approaches the shear plane (i.e. $\varsigma \to +\infty$). Additionally, the reduction in the minimum surface pressure magnitudes due to shear-thinning becomes more significant as $\varsigma \to +\infty$ and vice-versa. The relationship between the particles orbital minimum tip pressure $\bar{P}_{min}$ and the initial particles out-of-plane orientation $\theta^0$ appearing in Figure 22b clearly shows a gradual widening of the gap between the Newtonian and non-Newtonian pressure profiles.

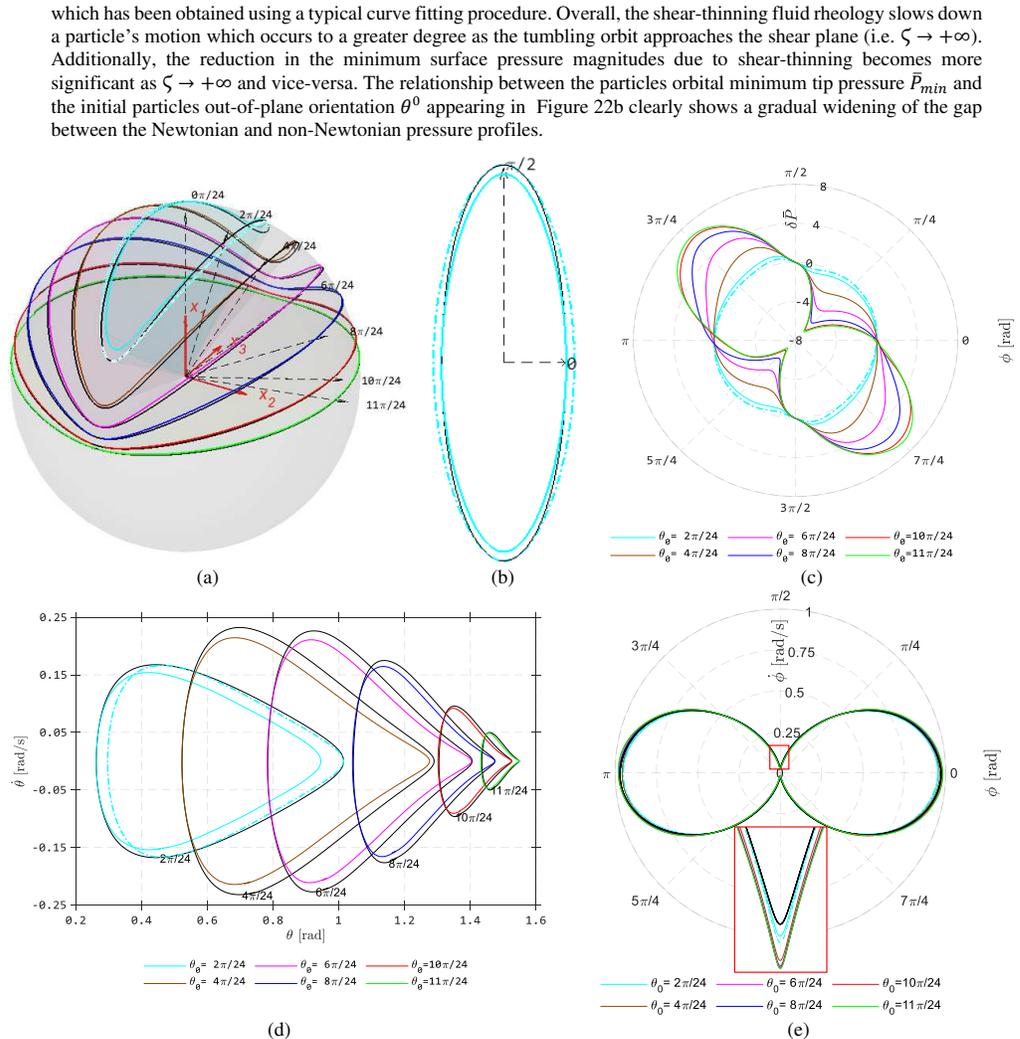

Figure 21: Effect of fluid shear-thinning on Jeffery's orbit: (a) particle's orbits in simple shear flow (b) dilated orbit (dashed cyan line $\phi^0 = \pi/2$), constricted orbit (continuous cyan line, $\phi^0 = 0$) and Newtonian orbit (black line) for $\theta^0 = 2\pi/24$ (c) difference in particle tip pressure between NT and GNF fluid (d) phase diagram of azimuth angle $\theta$ vs nutation $\dot{\theta}$ (e) polar plot of precession $\dot{\phi}$ vs polar angle $\phi$, for different initial particle orientation between $-2\pi/24 \leq \theta^0 \leq -12\pi/24$, $\phi^0 = 0$, $\psi^0 = 0$ and for NT fluid (dashed) and GNF power-law fluids (continuous) with $m = 1\ Pa \cdot s^n, n = 0.5$.

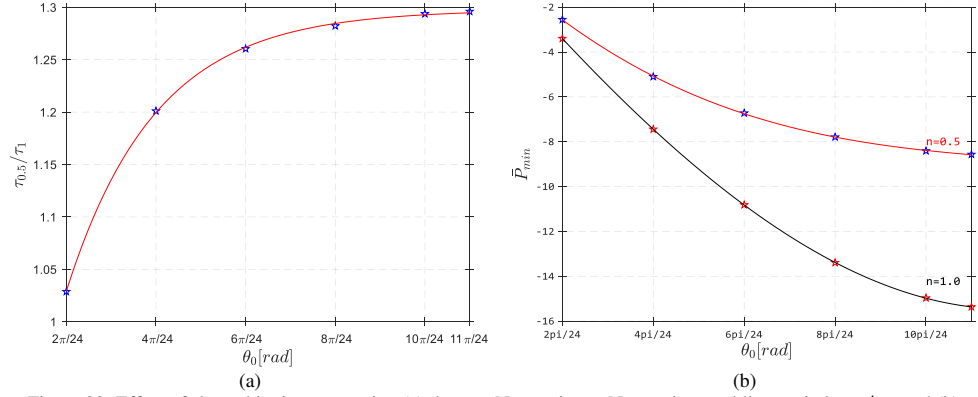

Figure 22: Effect of shear-thinning comparing (a) the non-Newtonian to Newtonian tumbling period $\tau_{0.5}/\tau_1$, and (b) the non-Newtonian (red line) and Newtonian (black line) particles orbital minimum tip pressure $\bar{P}_{min}$, versus the initial azimuth angle $\theta_0$, considering GNF power-law fluid, with with $m = 1\, Pa \cdot s^n, n = 0.5$ and initial orbit position $\phi^0 = 0$.

**<u>Effect of Geometric Aspect Ratio</u>**

Sensitivity study on the influence of the particle geometric aspect ratio on its field state showed that the aspect ratio significantly influences the observed particle kinematic behaviour and the surface pressure distribution in Newtonian shear flow. For completeness, we now consider the effect of the geometric aspect ratio on particle behaviour in shear-thinning simple shear flow making comparisons to the behaviour in a Newtonian fluid. Previous studies showed that the shear-thinning effect on the particle's orbit are magnified with increasing initial out of plane orientation $\theta^{0}$[34]. As such we consider Jeffery's orbit with initial particle orientation of $\phi^0 = 0, \theta^0 = 2\pi/24$, and $\psi^0 = 0$. Figure 23 shows the deviation in particle trajectories, pressure and dynamic responses between the shear-thinning and Newtonian fluid for various particle aspect ratios. For spherical shaped particles, shear-thinning has no significant effect on the particles orbit, or the evolution of the particle's surface pressure and dynamic responses. However, as the particle aspect ratio increases up to $r_e = 6$, we observe considerable deviation in the particle trajectory (cf. Figure 23a) consistent with the findings of Abtahi et al.[34]. Similar to results that appear above, the particle trajectory is elongated or constricted depending on the initial starting position on a particular Newtonian Jeffery's orbit. With a further increase in the particle's slenderness, i.e. as $\kappa \to 1$, modification of the particle's trajectory due to shear-thinning becomes negligible as was also observed by Ferec et al.[62].

The shear-thinning effect on the pressure response however continues to increase with the particle length (cf. Figure 23b) which can be attributed to the hydrostatic stress intensification at the particle's tip arising from the increased particle length and/or the related decrease in the tip curvature. Likewise, the impact of shear-thinning on particle angular velocities is initially observed to increase with increasing aspect ratio (cf. Figure 23c & d). The non-linear effects, however, gradually declines with further increase in ellipsoid's slenderness. The shear-thinning behaviour is observed to slightly decrease the particles orbit period with slight increase in the aspect ratio. Further increases in the particle's slenderness, however, results in the shear-thinning behaviour prolonging the tumbling period. At lower aspect ratios, the pressure drag which does not depend on the local viscosity dominates the hydrodynamic resistance, however, with longer particles, the skin friction drag becomes significant due to the increased surface area and change in apparent viscosity[25]. Since a decrease in the apparent viscosity is known to slow down particle motion, we experience longer tumbling periods with considerable increase in the particle aspect ratio (cf. Table VII).

| $r_e$ | 1 | 2 | 3 | 6 | 10 |
|---|---|---|---|---|---|
| $\tau_{0.5}/\tau_1$ | 1.000 | 0.966 | 0.951 | 1.027 | 1.250 |

Table VII: Ratio of the particle's tumbling period in shear-thinning simple shear flow with $n = 0.5$ and $m = 1\, Pa \cdot s^n$, and $\dot{\gamma} = 1s^{-1}$ and with to the reference Newtonian quantity, $\tau_1$. Initial particle orientation is $\phi^0 = 0, \theta^0 = 2\pi/24, \psi^0 = 0$.

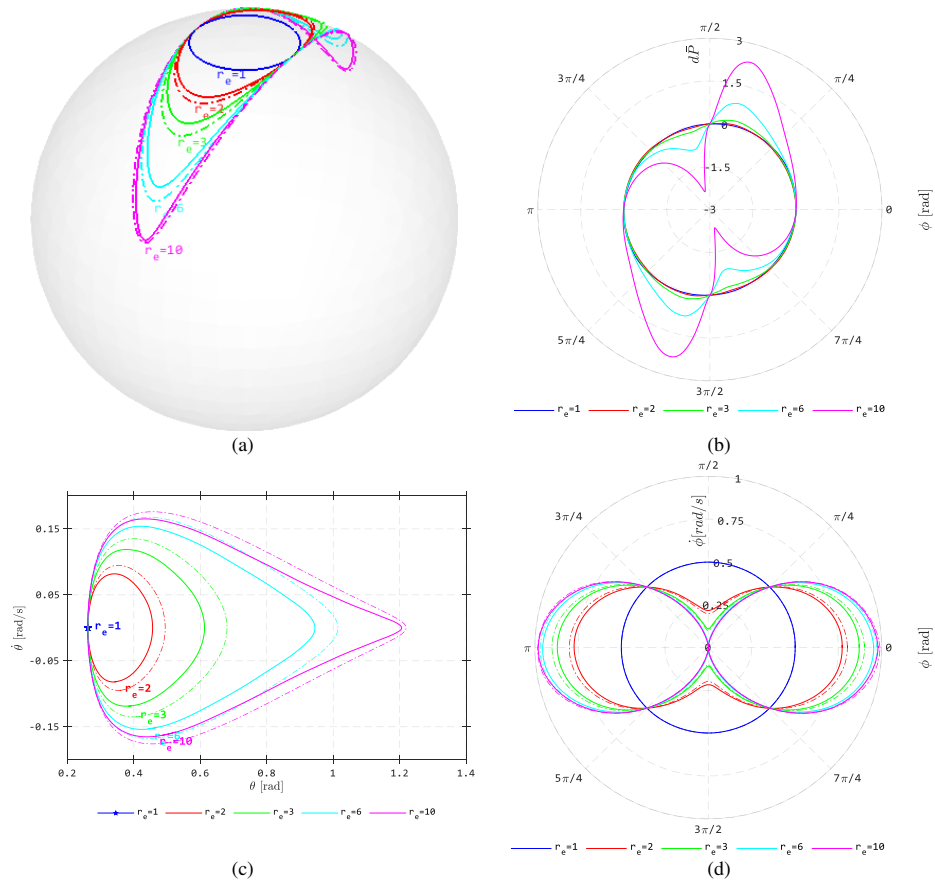

Figure 23: Showing (a) particle's orbits in simple shear flow (b) difference in particle tip pressure evolution between NT and GNF fluid (c) phase diagram of azimuth angle $\theta$ vs nutation $\dot{\theta}$ (d) polar plot of the polar angle $\phi$ vs precession $\dot{\phi}$, for different particle aspects $r_e$ and for NT fluid (dashed) and GNF power-law fluids (continuous) with $m = 1\ Pa \cdot s^n, n = 0.5$. Initial particle orientation is $\phi^0 = 0, \theta^0 = 2\pi/24$, $\psi^0 = 0$.

Since typical EDAM printed fiber-filled polymer composites are known to have very high aspect ratios $r_e > 45$[78,79], the shear-thinning rheology is expected to have negligible effects on particle angular velocity and trajectory. However, we expect the non-Newtonian fluid slows down the particles kinematics and reduces the surface pressure distribution.

**Conclusion**

In conclusion, a non-linear 3D-FEM numerical approach has been implemented to investigate the effects of shear-thinning fluid rheology in combination with other factors including the particles aspect ratio and initial particle orientation on the dynamics and surface pressure distribution on a particle suspended in viscous homogenous flow. The particles behavior in a special class of homogenous flows that typifies conditions found in melt flow regions of the of an extrusion nozzle during polymer composite additive manufacturing processing is also studied.

In the Newtonian flow, the ellipsoidal particle stalls in extension dominant asymmetric flow regimes but tumbles periodically in axisymmetric flows irrespective of the magnitude of the extension rate. The stall event in asymmetric flows is dictated by the shear-to-extension rate ratio. Increased shear dominance increases flow symmetry and tendency for continuous and periodic particle tumbling. The tumbling period in the asymmetric flows is expectedly dependent on the shear-to-extension rate ratio. The tumbling period increases asymptotically with increasing extension dominance until the conditions for stall based on Jeffery's equation are satisfied. On the other hand, the evolution time to particle stall is shown to increase asymptotically with increased shear dominance until the conditions for stall are violated. With sustained particle motion, the particle tip pressure fluctuates between extremums at the instants where its orientation aligns with the principal flow directions. An increase in the ellipsoidal particle aspect ratio was shown to affect the particles dynamics and increase the tumbling period. It also was shown to exacerbate the pressure extremes at the particle tip which could be caused by the increased aspect ratio alone, or the related reduction in tip curvature, or both. With a narrowing of Jeffery's orbit as the particle tumbles further out of plane, the particle surface pressure extremes are observed to decrease and the surface location of the pressure extreme further deviates from the particle's tip location. The orbital peak particle tip pressure magnitude follows a somewhat linear relationship with the polar location on the orbit across spectrum of degenerate Jeffery's orbit.

The behavior of the suspended particle is shown to be affected by the shear-thinning fluid rheology. In the axisymmetric flows where the particle motion ensues periodically, the shear-thinning fluid rheology slows down the particles motion and increases the tumbling period. Cessation of particle motion (i.e., a stall condition) in the asymmetric homogenous flows is shown to be dictated by a competing influence of the shear-thinning intensity and shear to extension rate dominance. The shear-thinning was found to have similar effect has decreasing the shear-rate dominance of the prevailing flow on the particles motion. Irrespective of the homogenous flow type, the magnitude of the particle surface pressure distribution was observed to significantly decrease with increased shear-thinning intensity due to an accompanying decrease in the effective viscosity of the fluid around the particle surface. The orbital location at which the pressure magnitude extremes on the particles surface are, however, unaffected by the shear-thinning rheology. On the shear-plane, shear-thinning rheology has no noticeable effect on the particles orbit, however, with a narrowing of the Jeffery orbit as we move further out of plane, the particle's trajectory deviates further from the Newtonian reference path. The shear-thinning rheology may either constrict or dilate the Newtonian orbit depending on the initial starting location of the particle on the orbit. The elongation of the particle's motion and the lowering of the pressure on the surface of the particle by the shear-thinning effect is augmented with widening of Jeffery's orbit as the particle tumbles closer to the shear plane. For spherical particles, the shear-thinning fluid has no significant effect on the dynamics or surface pressure distribution, but with increased aspect ratio, modification of the particle's trajectory and dynamics due to the non-linear effects becomes significant until a critical point, where the non-linear effects are reversed. With excessive particle slenderness, the impact of the shear-thinning fluid on the particle's trajectory and dynamics diminishes. On the contrary, the effects of the shear-thinning on lowering of the particle's surface pressure magnitude is proportionally elevated with increasing aspect ratio.

The foregoing discussion on the study of rigid spheroidal particle's behavior in viscous homogenous suspension are applicable in understanding and control various key transport phenomenon of this FSI process. For instance, the fluctuation of local field surface pressure distribution of suspended fibers in polymeric processes identified by Awenlimobor et al.[63] as a key mechanism potentially responsible for porosities within the composite beads could be controlled by suitable rheological adjustment to reduce the local pressure fluctuations. On one hand, increasing the shear-thinning intensity may help control the void formations, however increased shear-thinning may increase the likelihood of multiphase flow segregation within nozzle and the create more anisotropy in the microstructure of the printed composite. The present study contributes to understanding the combined effects of various flow parameters on the flow-field development during polymer composite processing which is instrumental in effectively controlling the expected microstructural behavior of printed composite beads and resulting properties.

**Acknowledgement:** The authors wish to acknowledge the National Science Foundation (NSF) for providing the necessary grants that made this publication possible [grant number: 2055628].

**Conflict of Interest Statement:** The authors have no conflicts to disclose.

# References

1. P. R. Vijayaratnam, C. C. O'Brien, J. A. Reizes, T. J. Barber, and E. R. Edelman, "The impact of blood rheology on drug transport in stented arteries: Steady simulations," PLoS One 10, e0128178 (2015).
2. Brenken, B., Barocio, E., Favaloro, A., Kunc, V. and Pipes, R.B., 2018. Fused filament fabrication of fiber-reinforced polymers: A review. Additive Manufacturing, 21, pp.1-16.
3. A. C. Barbati, J. Desroches, A. Robisson, and G. H. McKinley, "Complex fluids and hydraulic fracturing," Annu. Rev. Chem. Biomol. Eng. 7, 415 (2016).
4. Zhang, T., Wu, J. and Lin, X., 2020. Numerical investigation on formation and motion of bubble or droplet in quiescent flow. Physics of Fluids, 32(3).
5. Datt, C. and Elfring, G.J., 2018. Dynamics and rheology of particles in shear-thinning fluids. Journal of Non-Newtonian Fluid Mechanics, 262, pp.107-114.
6. Kugler, S.K., Kech, A., Cruz, C. and Osswald, T., 2020. Fiber orientation predictions—a review of existing models. *Journal of Composites Science*, *4*(2), p.69.
7. Evans, J. G. V., The flow of a suspension of force-free rigid rods in a Newtonian fluid, Ph.D. thesis, University of Cambridge, Cambridge, 1975.
8. Einstein, A., 1905. Eine neue bestimmung der moleküldimensionen (Doctoral dissertation, ETH Zurich).
9. Einstein, A., 1911. Berichtigung zu meiner arbeit: Eine neue bestimmung der moleküldimensionen. Annalen der Physik, 339(3), pp.591-592.
10. Mezi, D., Ausias, G., Advani, S.G. and Férec, J., 2019. Fiber suspension in 2D nonhomogeneous flow: The effects of flow/fiber coupling for Newtonian and power-law suspending fluids. Journal of Rheology, 63(3), pp.405-418.
11. R.G. Larson, The Structure and Rheology of Complex Fluids, Oxford University Press, New York, 1999.
12. Leal, L.G., 1980. Particle motions in a viscous fluid. Annual Review of Fluid Mechanics, 12(1), pp.435-476.
13. A. Oberbeck, "Ueber stationare Flüssigkeitsbewegungen mit Berücksichtigung der inneren reibung," J. Reine Angew. Math. 1876, 62.
14. D. Edwardes, "Steady motion of a viscous liquid in which an ellipsoid is constrained to rotate about a principal axis," Q. J. Pure Appl. Math. 26, 70 (1893).
15. Jeffery, G. B. The motion of ellipsoidal particles immersed in a viscous fluid. Proceedings of the Royal Society of London. Series A, 102(715):161-179, 1922.
16. H.L. Goldsmith, S.G. Mason, The micro rheology of dispersions, Rheology: Theory and Applications, vol. 4, Academic Press Inc., New York, 1967, pp. 85–250.
17. Bretherton, F.P., 1962. The motion of rigid particles in a shear flow at low Reynolds number. Journal of Fluid Mechanics, 14(2), pp.284-304.
18. Cox, R.G., 1965. The steady motion of a particle of arbitrary shape at small Reynolds numbers. Journal of Fluid Mechanics, 23(4), pp.625-643.
19. Taylor, G.I., 1923. The motion of ellipsoidal particles in a viscous fluid. Proceedings of the Royal Society of London. Series A, Containing Papers of a Mathematical and Physical Character, 103(720), pp.58-61.
20. Saffman, P.G., 1956. On the motion of small spheroidal particles in a viscous liquid. Journal of Fluid Mechanics, 1(5), pp.540-553.
21. Laurencin, T., Orgéas, L., Dumont, P.J., du Roscoat, S.R., Laure, P., Le Corre, S., Silva, L., Mokso, R. and Terrien, M., 2016. 3D real-time and in situ characterisation of fibre kinematics in dilute non-Newtonian fibre suspensions during confined and lubricated compression flow. Composites Science and Technology, 134, pp.258-266.
22. Geißler, P., Domurath, J., Ausias, G., Férec, J. and Saphiannikova, M., 2023. Viscosity and dynamics of rigid axisymmetric particles in power-law fluids. *Journal of Non-Newtonian Fluid Mechanics*, *311*, p.104963.
23. Gunes, D.Z., Scirocco, R., Mewis, J. and Vermant, J., 2008. Flow-induced orientation of non-spherical particles: Effect of aspect ratio and medium rheology. Journal of Non-Newtonian Fluid Mechanics, 155(1-2), pp.39-50.
24. D'Avino, G. and Maffettone, P.L., 2015. Particle dynamics in viscoelastic liquids. Journal of Non-Newtonian Fluid Mechanics, 215, pp.80-104. D'Avino, G. and Maffettone, P.L., 2015. Particle dynamics in viscoelastic liquids. Journal of Non-Newtonian Fluid Mechanics, 215, pp.80-104.
25. Tripathi, A., Chhabra, R.P. and Sundararajan, T., 1994. Power law fluid flow over spheroidal particles. *Industrial & engineering chemistry research*, *33*(2), pp.403-410.

26. Xu, X., Ren, H., Chen, S., Luo, X., Zhao, F. and Xiong, Y., 2023. Review on melt flow simulations for thermoplastics and their fiber reinforced composites in fused deposition modeling. *Journal of Manufacturing Processes*, *92*, pp.272-286.
27. Slattery, J.C., 1972. Momentum, energy, and mass transfer in continua. *(No Title)*.
28. Wasserman, M. L.; Slattery, J. C. Upper and lower bounds on the drag coefficient of a sphere in a power model fluid. AIChE J. 1964, 10, 383-388.
29. Chhabra, R.P.; Uhlherr, P. T. Creeping motion of spheres through shear-thinning elastic fluids described by the Carreau viscosity equation. Rheol. Acta 1980,19,187-195.
30. Hopke, S. W.; Slattery, J. C. Upper and lower bounds on the drag coefficient of a sphere in an Ellis model fluid. AIChE J. 1970,16, 224-229
31. Férec, J., Bertevas, E., Khoo, B.C., Ausias, G. and Phan-Thien, N., 2021. Rigid fiber motion in slightly non-Newtonian viscoelastic fluids. *Physics of Fluids*, *33*(10).
32. L.G. Leal, The slow motion of slender rod-like particles in a second-order fluid, J. Fluid Mech. 69 (1975) 305–337.
33. Brunn, P., 1977. The slow motion of a rigid particle in a second-order fluid. Journal of Fluid Mechanics, 82(3), pp.529-547.
34. Abtahi, S.A. and Elfring, G.J., 2019. Jeffery orbits in shear-thinning fluids. Physics of Fluids, 31(10), p.103106.
35. Liu, M.B. and Liu, G., 2010. Smoothed particle hydrodynamics (SPH): an overview and recent developments. *Archives of computational methods in engineering*, *17*, pp.25-76.
36. Li, S. and Liu, W.K., 2002. Meshfree and particle methods and their applications. *Appl. Mech. Rev.*, *55*(1), pp.1-34.
37. Joung, C.G., 2003. Direct simulation studies of suspended particles and fibre-filled suspensions.
38. He, L., Lu, G., Chen, D., Li, W. and Lu, C., 2017. Three-dimensional smoothed particle hydrodynamics simulation for injection molding flow of short fiber-reinforced polymer composites. *Modelling and Simulation in Materials Science and Engineering*, *25*(5), p.055007.
39. Bertevas, E., Férec, J., Khoo, B.C., Ausias, G. and Phan-Thien, N., 2018. Smoothed particle hydrodynamics (SPH) modeling of fiber orientation in a 3D printing process. *Physics of Fluids*, *30*(10).
40. Ouyang, Z., Bertevas, E., Parc, L., Khoo, B.C., Phan-Thien, N., Férec, J. and Ausias, G., 2019. A smoothed particle hydrodynamics simulation of fiber-filled composites in a non-isothermal three-dimensional printing process. *Physics of Fluids*, *31*(12).
41. S. Yashiro, H. Sasaki, and Y. Sakaida, "Particle simulation for predicting fiber motion in injection molding of short-fiber-reinforced composites," Composites, Part A 43, 1754–1764 (2012).
42. Yashiro, S., Okabe, T. and Matsushima, K., 2011. A numerical approach for injection molding of short-fiber-reinforced plastics using a particle method. *Advanced Composite Materials*, *20*(6), pp.503-517.
43. M. Ellero and R. I. Tanner, "SPH simulations of transient viscoelastic flows at low Reynolds number," J. Non-Newtonian Fluid Mech. 132, 61–72 (2005).
44. X. J. Fan and R. Z. R. I. Tanner, "Smoothed particle hydrodynamics simulation of non-Newtonian moulding flow," J. Non-Newtonian Fluid Mech. 165, 219–226 (2010).
45. Xu, X., Ouyang, J., Yang, B. and Liu, Z., 2013. SPH simulations of three-dimensional non-Newtonian free surface flows. *Computer Methods in Applied Mechanics and Engineering*, *256*, pp.101-116.
46. Xiang, H. and Chen, B., 2015. Simulating non-Newtonian flows with the moving particle semi-implicit method with an SPH kernel. *Fluid Dynamics Research*, *47*(1), p.015511.
47. Yamanoi, M.; Maia, J.M. Stokesian dynamics simulation of the role of hydrodynamic interactions on the behavior of a single particle suspending in a Newtonian fluid. Part 1. 1D flexible and rigid fibers. J. Non-Newton. Fluid Mech. 2011, 166, 457–468.
48. Andri´c, J.; Fredriksson, S.T.; Lindström, S.B.; Sasic, S.; Nilsson, H. A study of a flexible fiber model and its behavior in DNS of turbulent channel flow. Acta Mech. 2013, 224, 2359–2374.
49. Skjetne, P.; Ross, R.F.; Klingenberg, D.J. Simulation of single fiber dynamics. J. Chem. Phys. 1997, 107, 2108–2121.
50. Yamamoto, S.; Matsuoka, T. A method for dynamic simulation of rigid and flexible fibers in a flow field. J. Chem. Phys. 1993, 98, 644–650.

51. Yamamoto, S.; Matsuoka, T. Viscosity of dilute suspensions of rodlike particles: A numerical simulation method. J. Chem. Phys. 1994, 100, 3317–3324.
52. Sobhani, S.M.J., Bazargan, S. and Sadeghy, K., 2019. Sedimentation of an elliptic rigid particle in a yield-stress fluid: A lattice-Boltzmann simulation. Physics of Fluids, 31(8), p.081902.
53. Ji, J., Zhang, H., An, X. and Yang, D., 2024. Numerical study of the interaction between cylindrical particles and shear-thinning fluids in a linear shear flow. *Physics of Fluids*, *36*(8).
54. Bay, R.S. and Tucker III, C.L., 1992. Fiber orientation in simple injection moldings. Part I: Theory and numerical methods. *Polymer composites*, *13*(4), pp.317-331.
55. Bay, R.S. and Tucker III, C.L., 1992. Fiber orientation in simple injection moldings. Part II: Experimental results. *Polymer composites*, *13*(4), pp.332-341.
56. Mu, Y., Zhao, G., Chen, A. *et al.* Numerical investigation of three-dimensional fiber suspension flow by using finite volume method. *Polym. Bull.* **74**, 4393–4414 (2017).
57. N. Phan-Thien, T. Tran-Cong, and A. L. Graham, Shear Flow of Periodic Arrays of Particle Clusters: A Boundary-Element Method, Journal of Fluid Mechanics 228, 275 (1991).
58. X.-J. Fan, N. Phan-Thien, and R. Zheng, Complete Double Layer Boundary Element Method for Periodic Suspensions, Zeitschrift fur angewandte mathematik und Physik ZAMP 49, 167 (1998).
59. S. Nasseri, N. Phan-Thien, and X. Fan, Lubrication Approximation in Completed Double Layer Boundary Element Method, Computational Mechanics 26, 388 (2000)
60. Sugihara-Seki, M., 1996. The motion of an ellipsoid in tube flow at low Reynolds numbers. *Journal of Fluid Mechanics*, *324*, pp.287-308.
61. Zhang, D. and Smith, D.E., 2016. Dynamic simulation of discrete fiber motion in fiber-reinforced composite materials processing. *Journal of Composite Materials*, *50*(10), pp.1301-1319.
62. Férec, J., Ausias, G. and Natale, G., 2018, May. Numerical evaluation of a single ellipsoid motion in Newtonian and power-law fluids. In AIP Conference Proceedings (Vol. 1960, No. 1, p. 020006). AIP Publishing LLC.
63. Awenlimobor, A., Smith, D.E. and Wang, Z., 2024. Simulation of fiber-induced melt pressure fluctuations within large scale polymer composite deposition beads. *Additive Manufacturing*, *80*, p.103980.
64. Awenlimobor, A., Smith, D.E. and Wang, Z., 2023. Investigating the Effect of Generalized Newtonian Fluid on the Micro-Void Development within Large Scale Polymer Composite Deposition Beads.
65. J. Sun, M. D. Smith, R. C. Armstrong, and R. A. Brown, Finite Element Method for Viscoelastic Flows Based on the Discrete Adaptive Viscoelastic Stress Splitting and the Discontinuous Galerkin Method: 223 DAVSS-G/DG, Journal of Non-Newtonian Fluid Mechanics 86, 2 81 (1999).
66. Wang, Z., Fang, Z., Xie, Z. and Smith, D.E., 2022. A review on microstructural formations of discontinuous fiber-reinforced polymer composites prepared via material extrusion additive manufacturing: fiber orientation, fiber attrition, and micro-voids distribution. *Polymers*, *14*(22), p.4941.
67. Leaf, L.G. and Hinch, E.J., 1974. Theoretical studies of a suspension of rigid particles affected by Brownian couples. *Rheologica Acta*, *13*(4), pp.891-891.
68. Zhang, D. and Smith, D.E., 2015. Finite element-based brownian dynamics simulation of nanofiber suspensions using Monte Carlo Method. Journal of Micro and Nano-Manufacturing, 3(4).
69. Zhang, D., 2013. Flow-Induced Micro-and Nano-Fiber Suspensions in Short-Fiber Reinforced Composite Materials Processing. University of Missouri-Columbia.
70. Zhang, D. and Smith, D.E., 2016. Dynamic simulation of discrete fiber motion in fiber-reinforced composite materials processing. Journal of Composite Materials, 50(10), pp.1301-1319.
71. Kim, S., 1986. Singularity solutions for ellipsoids in low-Reynolds-number flows: with applications to the calculation of hydrodynamic interactions in suspensions of ellipsoids. *International journal of multiphase flow*, *12*(3), pp.469-491.
72. Cintra Jr, J.S. and Tucker III, C.L., 1995. Orthotropic closure approximations for flow-induced fiber orientation. Journal of Rheology, 39(6), pp.1095-1122.
73. Schuller, T., Fanzio, P. and Galindo-Rosales, F.J., 2023. The impact of polymer rheology on the extrusion flow in fused filament fabrication. arXiv preprint arXiv:2311.05158.
74. Giusteri, G.G. and Seto, R., 2018. A theoretical framework for steady state rheometry in generic flow conditions. Journal of Rheology, 62(3), pp.713-723.
37/40

ACCEPTED MANUSCRIPT

Physics of Fluids

This is the author's peer reviewed, accepted manuscript. However, the online version of record will be different from this version once it has been copyedited and typeset.
PLEASE CITE THIS ARTICLE AS DOI: 10.1063/5.0242953

AIP Publishing75.   Lubansky, A.S., Boger, D.V., Servais, C., Burbidge, A.S. and Cooper-White, J.J., 2007. An approximate solution to flow through a contraction for high Trouton ratio fluids. Journal of non-newtonian fluid mechanics, 144(2-3), pp.87-97.
76.   Folgar, F. and Tucker III, C.L., 1984. Orientation behavior of fibers in concentrated suspensions. Journal of Reinforced Plastics and Composites, 3(2), pp.98-119.
77.   Reddy, J.N., 2014. An Introduction to Nonlinear Finite Element Analysis Second Edition: with applications to heat transfer, fluid mechanics, and solid mechanics. OUP Oxford.
78.   Wang, Z., Smith, D.E. and Jack, D.A., 2021. A statistical homogenization approach for incorporating fiber aspect ratio distribution in large area polymer composite deposition additive manufacturing property predictions. *Additive Manufacturing*, *43*, p.102006
79.   Russell, T. and Jack, D.A., 2019. Fiber Aspect Ratio Characterization and Stiffness Prediction in Large-Area, Additive Manufactured, Short-Fiber Composites. *Proceedings of the SPE ACCE*.
80.   Dinh, S.M. and Armstrong, R.C., 1984. A rheological equation of state for semi-concentrated fiber suspensions. Journal of Rheology, 28(3), pp.207-227.**Appendix A.     Definition of Constants in Jeffery's Equation**

The expressions of the components of the variable vector $\underline{\chi}$ and coefficient tensors $\underline{\Lambda}^I$, $\underline{\underline{\Lambda}}^{II}$ & $\underline{\underline{\Lambda}}^{III}$ that appear in the definition of the Jeffery's velocity and pressure are defined in A1-A6 below. For the variable vector $\underline{\chi}$ the components are given as

$$\chi_1 = \alpha' X_2 X_3, \qquad \chi_2 = \beta' X_3 X_1, \qquad \chi_3 = \gamma' X_1 X_2 \qquad \text{A1}$$

The components in $\underline{\Lambda}^I$ vector are likewise given as

$$R = -\frac{\tilde{f}}{\alpha'_0}, \qquad S = -\frac{\tilde{g}}{\beta'_0}, \qquad T = -\frac{\tilde{h}}{\gamma'_0} \qquad \text{A2}$$

The components in $\underline{\underline{\Lambda}}^{II}$ tensor are given as

$$U = 2[b^2 B - c^2 C], \qquad V = 2[c^2 C - a^2 A], \qquad W = 2[a^2 A - b^2 B] \qquad \text{A3}$$

where the coefficients $A, B, C$ in equation A3 above are also components of tensor $\underline{\underline{\Lambda}}^{III}$ containing the stresslet and torque acting on the rigid ellipsoidal particle suspended in linear ambient flow-field[71]. given in equation A4 below

$$
\begin{aligned}
A &= \frac{1}{6}\left\{\frac{2\alpha_0'' \tilde{a} - \beta_0'' \tilde{b} - \gamma_0'' \tilde{c}}{\beta_0'' \gamma_0'' + \gamma_0'' \alpha_0'' + \alpha_0'' \beta_0''}\right\}, & F &= \frac{\beta_0 \tilde{f} - c^2 \alpha_0'(\xi - \Psi_1)}{2\alpha_0'(b^2 \beta_0 + c^2 \gamma_0)}, & F' &= \frac{\gamma_0 \tilde{f} + b^2 \alpha_0'(\xi - \Psi_1)}{2\alpha_0'(b^2 \beta_0 + c^2 \gamma_0)} \\
B &= \frac{1}{6}\left\{\frac{2\beta_0'' \tilde{b} - \gamma_0'' \tilde{c} - \alpha_0'' \tilde{a}}{\beta_0'' \gamma_0'' + \gamma_0'' \alpha_0'' + \alpha_0'' \beta_0''}\right\}, & G &= \frac{\gamma_0 \tilde{g} - a^2 \beta_0'(\eta - \Psi_2)}{2\beta_0'(c^2 \gamma_0 + a^2 \alpha_0)}, & G' &= \frac{\alpha_0 \tilde{g} + c^2 \beta_0'(\eta - \Psi_2)}{2\beta_0'(c^2 \gamma_0 + a^2 \alpha_0)} \\
C &= \frac{1}{6}\left\{\frac{2\gamma_0'' \tilde{c} - \alpha_0'' \tilde{a} - \beta_0'' \tilde{b}}{\beta_0'' \gamma_0'' + \gamma_0'' \alpha_0'' + \alpha_0'' \beta_0''}\right\}, & H &= \frac{\alpha_0 \tilde{h} - b^2 \gamma_0'(\zeta - \Psi_3)}{2\gamma_0'(a^2 \alpha_0 + b^2 \beta_0)}, & H' &= \frac{\beta_0 \tilde{h} + a^2 \gamma_0'(\zeta - \Psi_3)}{2\gamma_0'(a^2 \alpha_0 + b^2 \beta_0)}
\end{aligned}
\qquad \text{A4}
$$

The Greek integral constants $\alpha$, $\beta$, $\gamma$ and their symmetric forms are defined as[15]

$$
\begin{aligned}
\alpha &= \int_\lambda^\infty \frac{1}{\Delta} \frac{d\lambda}{a^2 + \lambda}, & \alpha' &= \int_\lambda^\infty \frac{1}{\Delta} \frac{d\lambda}{(b^2 + \lambda)(c^2 + \lambda)}, & \alpha'' &= \int_\lambda^\infty \frac{1}{\Delta} \frac{\lambda d\lambda}{(b^2 + \lambda)(c^2 + \lambda)} \\
\beta &= \int_\lambda^\infty \frac{1}{\Delta} \frac{d\lambda}{b^2 + \lambda}, & \beta' &= \int_\lambda^\infty \frac{1}{\Delta} \frac{d\lambda}{(c^2 + \lambda)(a^2 + \lambda)}, & \beta'' &= \int_\lambda^\infty \frac{1}{\Delta} \frac{\lambda d\lambda}{(c^2 + \lambda)(a^2 + \lambda)} \\
\gamma &= \int_\lambda^\infty \frac{1}{\Delta} \frac{d\lambda}{c^2 + \lambda}, & \gamma' &= \int_\lambda^\infty \frac{1}{\Delta} \frac{d\lambda}{(a^2 + \lambda)(b^2 + \lambda)}, & \gamma'' &= \int_\lambda^\infty \frac{1}{\Delta} \frac{\lambda d\lambda}{(a^2 + \lambda)(b^2 + \lambda)}
\end{aligned}
\qquad \text{A5}
$$

A greek constant subscripted with 0, implies that the lower limit of integration $\lambda = 0$. The variables $\tilde{a}, \tilde{b}, \tilde{c}, \tilde{f}, \tilde{g}, \tilde{h}$ are components of the rate of deformation tensor $\underline{\underline{\Gamma}}$ and $\xi, \eta, \zeta$ are components of the vorticity tensor $\underline{\underline{\Xi}}$ i.e.

38/40

$$\underline{\underline{\Gamma}} = \begin{bmatrix} \tilde{a} & \tilde{h} & \tilde{g} \\ \tilde{h} & \tilde{b} & \tilde{f} \\ \tilde{g} & \tilde{f} & \tilde{c} \end{bmatrix}, \quad \underline{\underline{\Xi}} = \underline{\Xi} \times \underline{I} = \begin{bmatrix} & -\zeta & +\eta \\ +\zeta & & -\xi \\ -\eta & +\xi & \end{bmatrix}, \quad \underline{\Xi} = \begin{bmatrix} \xi \\ \eta \\ \zeta \end{bmatrix} \qquad \text{A6}$$

**Appendix B.    Flow-Regimes in Typical EDAM Nozzle**

Polymer composite melt flow through the nozzle in typical EDAM polymer composite processing is characterized by complex combination of shear and extensional deformation rate components that are dependent on the viscoelastic polymer melt rheology and the geometry of the extrusion nozzle. The flow condition at the nozzle wall is pure shear and at the nozzle centreline is pure uniaxial elongation (cf. Figure 24)[73,74]. Away from the convergent zone in the lubrication zone defined by the clearance between the screw edge and the nozzle walls, the flow is predominantly shear dominant while close to the centreline and near the entrance of the nozzle where the flow undergoes acceleration due to geometric constriction, the flow is dominated by extensional rate, and at the vortices created near the notch edges with sharp transitions due to elastic instabilities, the flow is mainly rotational[73]. The flow contraction region consists of a complex combination of the various flow categories with varying dominance.

A simple metric used to classify the flow regimes is based on a flow parameter $\bar{v}$ given by[73].

$$\bar{v} = \frac{\dot{\gamma}_c + j\omega_c}{\dot{\gamma}_c - j\omega_c} \qquad \text{B1}$$

where $\dot{\gamma}_c$ is the magnitude of deformation rate tensors defined as $\dot{\gamma}_c = \sqrt{2\Gamma_{ij}\Gamma_{ji}}$ and $\omega_c$ is the magnitude of the vorticity tensor given as $\omega_c = \sqrt{2\,\Xi_{ij}\,\Xi_{ji}}$. The flow is pure shear when $\bar{v} = 0$, pure elongational when $\bar{v} = 1$, and purely rotational when $\bar{v} = -1$. Typical flow patterns within the convergent zone results in $\bar{v}$ lying between $-1 \leq \bar{v} \leq 1$.

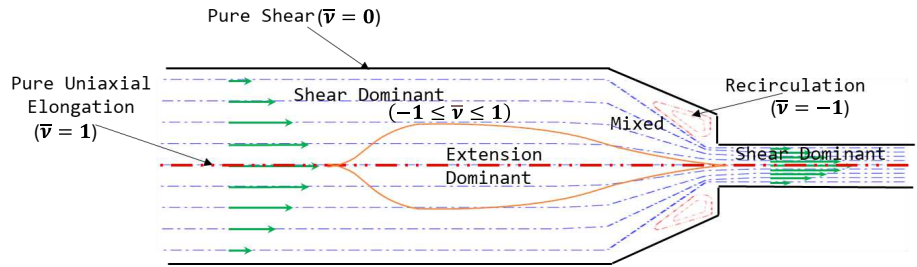

Figure 24: Schematic showing flow regimes within a typical EDAM nozzle during polymer processing.

**Appendix C.    Obtaining Particle Stall Orientation Angles in Newtonian Homogenous Flows**

The particle stall angles under favorable conditions in general class of homogenous flows can be obtained using the tensorial representation for the particle orientation of an axisymmetric ellipsoidal particle in viscous suspension with velocity gradient $\underline{\underline{L}}$ developed by Dinh et al.[80] based on Jeffery's model assumptions and is given as

$$\dot{\rho}_i = -\Xi_{ij}\rho_j + \kappa\big(\Gamma_{ij}\rho_j - \Gamma_{kl}\rho_k\rho_l\rho_i\big) \qquad \text{C1}$$

where $\underline{\rho}$ is the particle orientation defined by the vector:

$$\underline{\rho} = [\cos\theta \quad \sin\theta\sin\phi \quad \sin\theta\cos\phi]^T \qquad \text{C2}$$

The Euler angles and angular velocities can be backtracked from the rate of the orientation vectors $\underline{\dot{\rho}}$ thus:

$$\phi = \tan^{-1}\frac{\rho_2}{\rho_3}, \qquad \theta = \cos^{-1}\rho_1, \qquad \dot{\phi} = \frac{\dot{\rho}_3}{\rho_3}\left[\frac{\dot{\rho}_2}{\dot{\rho}_3} - \frac{\rho_2}{\rho_3}\right]\left[1 + \frac{\rho_2^2}{\rho_3}\right]^{-1}, \qquad \dot{\theta} = -\dot{\rho}_1(1-\rho_1^2)^{-1/2} \qquad \text{C3}$$

Considering the normalization condition, the independent components of the particle orientation at stall can likewise be obtained via the Newton-Raphson numerical iterative process according to eqn. xx. below

$$\underline{\rho}_s^+ = \underline{\rho}_s^- - \underline{J}_{\theta_2}^{-1}\underline{\dot{\theta}}^\rho \qquad \text{C4}$$

where $\underline{\rho}_s = [\rho_2^s \quad \rho_3^s]^T$, $\sum_{\forall j}\rho_j = 1$, $\underline{\dot{\theta}}^\rho = [\dot{\phi} \quad \dot{\theta}]^T$, and the components of the Jacobian $\underline{J}_{\theta_2}$ are explicitly defined in eqns. C5-C8 below

$$\underline{J}_{\theta_2,11} = \frac{1}{\rho_3}\left\{J_{\rho,21} - J_{\rho,31}\frac{\rho_2}{\rho_3} - \frac{\dot{\rho}_3}{\rho_3}\left[1 + 2\left(\frac{\dot{\rho}_2}{\dot{\rho}_3} - \frac{\rho_2}{\rho_3}\right)\left(\frac{\rho_2}{\rho_3} + \frac{\rho_3}{\rho_2}\right)^{-1}\right]\right\}\left[1+\frac{\rho_2^2}{\rho_3}\right]^{-1} \qquad \text{C5}$$

$$\underline{J}_{\theta_2,12} = \frac{1}{\rho_3}\left\{J_{\rho,22} - J_{\rho,32}\frac{\rho_2}{\rho_3} - \frac{\dot{\rho}_3}{\rho_3}\frac{\dot{\rho}_2}{\dot{\rho}_3} + 2\frac{\dot{\rho}_3}{\rho_3}\frac{\rho_2}{\rho_3}\left[1 + \left(\frac{\dot{\rho}_2}{\dot{\rho}_3} - \frac{\rho_2}{\rho_3}\right)\left(\frac{\rho_2}{\rho_3} + \frac{\rho_3}{\rho_2}\right)^{-1}\right]\right\}\left[1+\frac{\rho_2^2}{\rho_3}\right]^{-1} \qquad \text{C6}$$

$$\underline{J}_{\theta_2,21} = [-J_{\rho,11}(1-\rho_1^2) + \dot{\rho}_1](1-\rho_1^2)^{-3/2} \qquad \text{C7}$$

$$\underline{J}_{\theta_2,22} = [-J_{\rho,12}(1-\rho_1^2) + \dot{\rho}_1](1-\rho_1^2)^{-3/2} \qquad \text{C8}$$

And the tensor $\underline{J}_\rho$ is computed from eqn. C9 below

$$\underline{J}_\rho = -\underline{\underline{\Xi}}\,\underline{\Omega} + \kappa\left[\underline{\underline{\Gamma}} - \underline{\rho}\underline{\rho}^T\left(\underline{\underline{\Gamma}} + \underline{\underline{\Gamma}}^T\right) - \left(\underline{\rho}^T\underline{\underline{\Gamma}}\underline{\rho}\right)\underline{\underline{I}}\right]\underline{\Omega}, \qquad \underline{\Omega}^T = \begin{bmatrix} -1 & 1 & 0 \\ -1 & 0 & 1 \end{bmatrix} \qquad \text{C9}$$

**Appendix D. Principal Flow Directions**

The principal flow directions can be obtained by spectral decomposition of the symmetric part of the velocity gradient tensor $\underline{\underline{\Gamma}}$. The respective eigenvectors $\underline{\Phi}^k$ are the principal flow directions. i.e.

$$\underline{\Phi}\,\big|\,\Gamma_{mn} = \Phi_{mk}\Lambda_{kl}\Phi_{nl}, \qquad \Lambda_{kl} = \begin{cases}\lambda_k & k = l \\ 0 & k \neq l\end{cases}, \qquad \underline{\Phi}\,\big|\,\underline{\underline{\Gamma}} = \underline{\underline{\Phi}}\,\underline{\underline{\Lambda}}\,\underline{\underline{\Phi}}^T \qquad \text{D1}$$

Considering the in-plane homogenous flow velocity gradient of Equation 22, the principal flow directions in the shear plane irrespective of coordinate reference frame are obtained as

$$\tan\phi_p = \frac{\dot{\varepsilon}_2 - \dot{\varepsilon}_3}{\dot{\gamma}} \pm \sqrt{\frac{\dot{\varepsilon}_2 - \dot{\varepsilon}_3}{\dot{\gamma}}^2 + 1} \qquad \text{D2}$$

As would be seen from the simulation results, for a particle tumbling in the flow shear-plane, the particle orientation at the location of minimum pressure extreme on particle's surface corresponds to position of particle alignment with one of the principal flow directions in the flow shear plane. i.e.

$$\underline{\rho}\Big|_{p=p_{max}} = \underline{\Phi}^k \quad \theta = 0 \qquad \text{D3}$$

Hence the peak pressure occurs at an instant $t_p$ such that $\phi(t_p) = \phi_p$.